\documentclass[%
 twocolumn,
 superscriptaddress,
 amsmath,amssymb,
 aps,prl,
citeautoscript,
]{revtex4-1}

\usepackage{graphicx}% Include figure files
\usepackage{dcolumn}% Align table columns on decimal point
\usepackage{bm}% bold math

\usepackage[colorlinks = true,
            linkcolor = blue,
            urlcolor  = blue,
            citecolor = blue,
            anchorcolor = blue]{hyperref}
            
\usepackage{fdsymbol}
\usepackage{siunitx}

\begin{document}

\title{Magnetic coupling in colloidal clusters  for hierarchical self-assembly}

\author{Joe G. \surname{Donaldson}}
\affiliation{Department of Chemical Engineering, Delft University of Technology, 2629 HZ Delft, The Netherlands}
\affiliation{Current Address: Unilever R\&D, Colworth MK44 1LQ, UK}
\author{Peter \surname{Schall}}
\affiliation{Institute of Physics, University of Amsterdam, 1098XH Amsterdam, The Netherlands.}
\author{Laura\ \surname{Rossi}}\email{l.rossi@tudelft.nl}
\affiliation{Department of Chemical Engineering, Delft University of Technology, 2629 HZ Delft, The Netherlands}

\date{\today}

\begin{abstract}

Manipulating the way in which colloidal particles self-organise is a central challenge in the design of functional soft materials.
Meeting this challenge requires the use of building blocks that interact with one another in a highly specific manner.
Their fabrication, however, is limited by the complexity of the available synthesis procedures.
Here, we demonstrate that, starting from experimentally available magnetic colloids, we can create a variety of complex building blocks suitable for hierarchical self-organisation using a simple scalable process.
Using computer simulations, we compress spherical and cubic magnetic colloids in spherical confinement, and   investigate their suitability to form small clusters with reproducible structural and magnetic properties.
We find that, while the structure of these clusters is highly reproducible, their magnetic character depends on the particle shape. Only spherical particles have the rotational degrees of freedom to produce consistent magnetic configurations, whereas cubic particles frustrate the minimisation of the cluster energy, resulting in various magnetic configurations.
To highlight their potential for self-assembly, we demonstrate that already clusters of three magnetic particles form highly nontrivial Archimedean lattices, namely staggered kagome, bounce and honeycomb, when viewing different aspects of the same monolayer structure. The work presented here offers a conceptually different way to design materials by utilizing pre-assembled magnetic building blocks that can readily self-organise into complex structures. 

\end{abstract} 

\maketitle

A contemporary goal common in the soft matter field aims at creating building blocks with specific functionalities. Using these nano- to micro-scale building blocks, scientists are envisaging of engineering materials with controllable properties \cite{Ozin:2009cc,Soukoulis:2011jj,Maldovan:2006dh,Veselago:2006eo,Gardner:2011bz}. For this reason, recent years have seen the development of a plethora of approaches to colloidal particle preparation, from classical wet-chemistry synthesis methods \cite{Sacanna:2010bv,Sacanna:2012ku,Wang:2013hh,Kraft:2010uq,Crassous:2012dg,Youssef:2016kb,Zheng:2017ce,Bae:2015cq,Zhang:2015fe,Kim:2007cs} to physical and lithographic techniques \cite{Hoover:1990vq,Hernandez:2007gn,Zhao:2007ec,Badaire:2007ga,Lipomi:2012jh,Ni:2016ig,Pawar:2008kp}. Solely using a building block's shape is a powerful way to control structure formation \cite{Nguyen:2011jg,Damasceno:2012gi,Rossi:2015jf}, however to obtain increasingly functional building blocks, chemists have to imbibe them with a "code" that specifically defines the way in which the particles will spontaneously assemble. These "codes" are usually formulated by using chemical \cite{Wang:2012gda,Chen:2011be,DiazA:2020kq} or physical\cite{Kraft:2012ga,Sacanna:2010bv} surface modifications. 
While unconventional colloidal preparation methods are on the rise, synthetic complexity and low yields still remain the most common limiting factors to obtain complex macroscopic materials \textit{via} colloidal self-organisation. 
Recently, it has been shown that carefully designed preassembly of simple colloidal particles, with interactions programmed by DNA coatings, allows the preparation of a variety of crystalline structures\cite{Ducrot:2017cs}. 
Preassembly of readily available colloidal particles into defined structures that can be used themselves as building blocks, is a powerful method and allows the use of well-known traditional colloidal units to make exotic architectures. 
In this context, magnetic particles are promising candidates to tailor particle assembly\cite{Rossi:2018ef}. The main advantage is that magnetic dipolar interactions not only allow the direct formation of predefined structures without a supplemental need for chemical or physical functionalization, but also have the potential to enable tuning of the formed structure with the application of external magnetic fields\cite{Sacanna:2012ku,Rossi:2018ef}.
Here we explore, using computer simulation, the design of complex magnetic building blocks from  experimentally accessible magnetic particles. The magnetic spherical and cubic particles are compressed in spherical confinement, to form building blocks, mimicking known emulsion templating techniques\cite{Cho:2005cp}.
Depending on the starting number of particles in the compression environment, we obtain clusters composed of $n=(2-10)$ particles. 
We have elucidated both the structural and  magnetic configurations of the particles within the clusters. 
In our analysis we find that, while the structural organisation of the obtained building block is robust in both cases, the magnetic configuration is consistent for spheres but not cubes due to the intrinsic difference in anisotropy, which in the latter case causes frustration in the alignment of the dipoles. 
An observation suggesting that spheres, in this scenario, are better candidates for use in self-assembly studies. 
To conceptualise this assertion we show that clusters of three magnetic spheres have the ability to readily form extended assemblies in a hierarchical fashion. 
This work introduces a general principle with associated rules to experimentally design magnetic building blocks capable of self-organising into structures unattainable for the simple constituent magnetic colloids. 

\section*{Results and Discussion}

\begin{figure}
	\centering
	\includegraphics[width=0.5\textwidth]{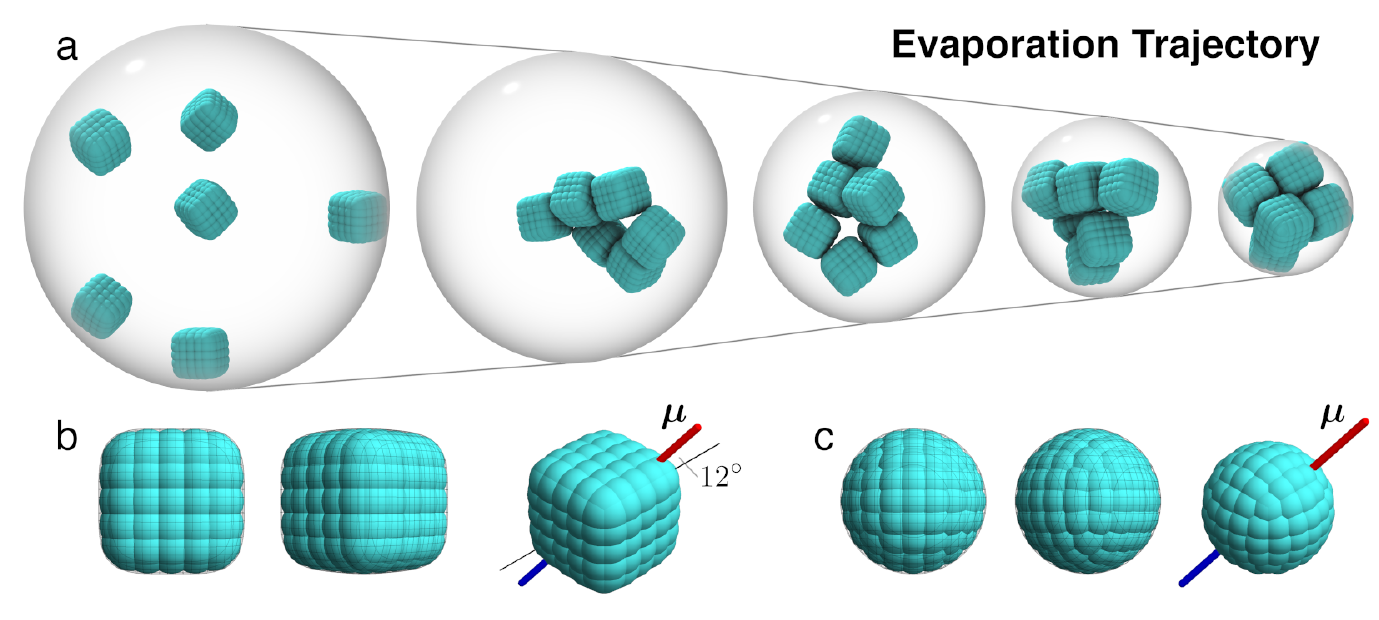}	
	\caption{{\bf Compression Mechanism and Particle Model}.
	{\bf(a)} A fixed number of particles was placed within a spherical confinement representing the emulsion droplet, an example for $n=6$ cubes is shown. The available volume is slowly decreased over the course of a simulation, resembling the evaporation of water from a droplet.
	{\bf(b)} Cubes are constructed from sub-units of spheres arranged to form the surface a superball geometry with a shape parameter of $m=4$. The wire-frame shown in the first two views is provided to highlight the exact superball surface. In the final view, the orientation of the particle dipole moment $\boldsymbol{\mu}$, is visualised with its $12^{\circ}$ tilt from the space diagonal. 
	{\bf(c)} Spheres, with a shape parameter of $m=2$, are constructed in an analogous fashion to facilitate comparisons. The approximation to perfect spherical geometry is indicated, again by the wire-frame. The orientation of $\boldsymbol{\mu}$ with respect to the particle geometry is no longer relevant due to symmetry, but indicated for completeness.
	}
\label{fig1}
\end{figure}

%=======================
% Compression Mechanism
%=======================
\noindent \textbf{Compression Mechanism}
\\
\\
The compression mechanism used to prepare clusters of magnetic particles is schematically shown in Figure~\ref{fig1}a. In order to emulate existing experimental procedures\cite{Cho:2005cp}, a fixed number $n$ of magnetic particles is placed randomly within a spherical confinement, initialised to a radius large enough to prevent the imposition of any confinement effects on the initial aggregation of the particles. The spherical volume is uniformly decreased over the course of a simulation to resemble the experimental observations during evaporation of emulsion droplets in colloidal cluster formation from water-in-oil emulsions\cite{Cho:2005cp}. For each cluster size we repeat the compression a total of 50 times, to test the reproducibility of the procedure and allow for the resulting structures to be compared\cite{Teich:2016ij}. In the simulations, the particles are propagated using Langevin molecular dynamics performed for a fixed number of particles, at a fixed temperature, and at a systematically varied 'fixed' volume. 
The scheme by which the droplet volume is reduced is discussed in the Methods section and visualised in Figure~\ref{fig6}, the protocol outlined follows an exponential decay to allow time for cluster equilibration as the droplet shrinks.
The interaction between particles consists of a short-range repulsion to prevent particle overlap, and the dipolar potential to characterise the long-range magnetic interaction. In experiments, clusters are formed when all the solvent in the droplet evaporates and the constituent particles are held together by van der Waals forces which arise upon particle contact. In simulation, each replica is considered complete when a force threshold is reached, indicating imminent confinement violation. Note that in our simulations we do not explicitly consider capillary forces, as these seem to be inconsequential in the formation of comparable colloidal clusters\cite{Peng:2013bb}. Similarly, due to the likelihood of low Reynolds number flow within individual droplets and the low density of the solvent, the hydrodynamic coupling between particles is expected to be slight and is thus neglected. Further information regarding the simulation protocol used is detailed in the Methods section and should be consulted prior to the subsequent sections to contextualise these results.
\\
\\

% ==== Figure 1 ====

%===============
% Cube Clusters
%===============
\noindent \textbf{Cube Clusters}
\\
% ==== Figure 2 ====
\begin{figure*}
	\centering
	\includegraphics[width=0.95\textwidth]{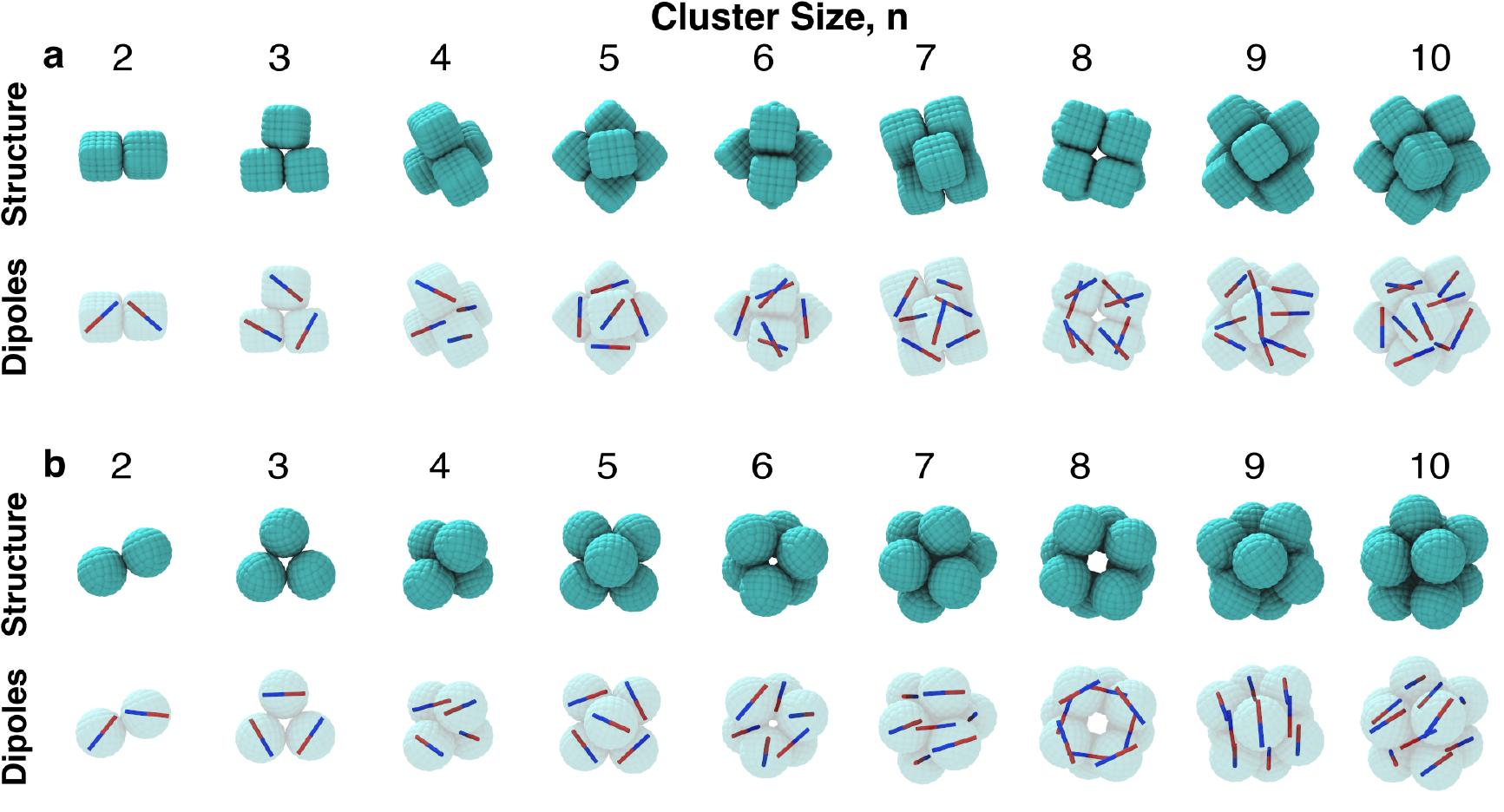}
	\caption{{\bf Clusters Post Confinement.} Visualisations of clusters for $n=(2-10)$ for (a) cubes and (b) spheres, achieved after the confinement procedure. The clusters shown represent the structures with the lowest 2\textsuperscript{nd} moment of the mass distribution, $\mathcal{M}_{2}$. The upper row of images in each figure shows the structure of the clusters obtained, this closely follows the progression of the platonic solids. The lower row gives a description of the magnetic character of the clusters. The dipole of individual particles is shown as a red and blue bar.
	}
    \label{fig2}
\end{figure*}

\noindent Cubes clusters are prepared using particles with rounded edges, a well-known feature of hematite colloids, the only known naturally occurring permanently magnetised micron-size colloidal system \cite{MartinezPedrero:2016gy,Lee:2009cy,Lee:2009km,Meijer:2013kb,Rossi:2018ef}. The choice of cubic-like particles follows from their precise anisotropic shape combined with well understood magnetic properties as reported by some of the authors in another work\cite{Rossi:2018ef}. The particles used for the simulations are illustrated in Figure~\ref{fig1}b. 
Their surface is constructed by overlapping spheres of equal diameter, these sub-units are arranged according to a superball geometry (Methods, Equation~\ref{eqn:superball}) with a shape parameter of $m=4$. While other shape parameters are undoubtedly of some interest, we chose to tailor our simulation to be as representative as possible of the physically accessible systems. The dipole moment $\boldsymbol{\mu}$ in such particles is known to lie at a face-tilted angle of $12^{\circ}$ from the space diagonal \cite{Rossi:2018ef}. The magnitude of $\boldsymbol{\mu}$ is set \textit{via} the experimentally derived dipole coupling parameter $\lambda$, the specifics of which can be found, along with further details regarding the particle model, in the Methods section. An overview of the clusters obtained for $n=(2-10)$ is displayed in Figure~\ref{fig2}a, where both the structure and dipolar configuration of representative clusters are reported. The clusters presented here are those with the lowest values of the 2\textsuperscript{nd} moment of the mass distribution $\mathcal{M}_{2}$. 

The top row of Figure~\ref{fig2}a highlights the arrangement of particles within the cluster, which is commensurate to that of non-magnetic spherical clusters as reported in both experiments\cite{Cho:2005cp,Manoharan:2003hb} and simulations\cite{Teich:2016ij,Lauga:2004im}. Small deviations in geometry are due to specific particle surface properties that can either promote particle adsorption to the interface\cite{Manoharan:2003hb}, or complete dispersion in the drying droplets\cite{Cho:2005cp}.  
This observation suggests that the magnetic interaction plays a secondary role to the confining forces. One can therefore expect that confinement is the driving force during evaporation. Turning to the lower row we show how the dipoles are configured within the clusters. Immediately we can see that the arrangement of the magnetic moments of the particles is frustrated, as can be seen by the absence of closed rings that are necessary to minimize the magnetic energy. It appears that a cube's sole route to minimise the magnetic flux of a given cluster is through the formation of approximate (quasi) anti-parallel pairs. As a result of this behaviour, the remanent magnetisation for cube clusters is often determined by a single particle  forced into an unfavourable magnetic configuration due to steric hinderance, this is most clearly seen in clusters for $n = 3$ and $5$.
\\
\\
%===============
% Sphere Clusters
%===============
\noindent \textbf{Sphere Clusters}
\\

\noindent Spherical particles with well-defined magnetization in the micron-size range are not easy to prepare from naturally occurring magnetic materials. This is because of the crystalline nature of most magnetic materials in combination with their general tendency to become multidomain at the sub-micron length scale.
However, it has recently been demonstrated that one can encase hematite cubes in a spherical polymeric shell \cite{Hueckel:2018is}, effectively producing spherical particles with a permanent dipole moment.
Accounting for the availability of this experimental protocol, we consider here the use of spherical particles that possess the same magnetic properties as the hematite cubes. We model our spherical particles in a fashion analogous to the cubes, in which sub-units of spheres are arranged according to a spherical geometry (superball $m=2$, Equation~\ref{eqn:superball}) with the same repulsive and dipole potentials active. Due to the re-introduction of spherical particle symmetry, the dipole moment orientation relative to the geometry is no longer relevant. The magnitude of the dipole moment and volume is kept constant between the particle types, given that these quantities are directly proportional. This procedure acts to realise an experimental version of hematite cube particles embedded in a spherical shell with diameter equal to the cube space diagonal. This equivalency is elaborated on further in the Methods section. An overview of the clusters obtained for $n=(2-10)$ is displayed in Figure~\ref{fig2}b, where, as before, the structure and dipolar configuration of the representative clusters are shown in the upper and lower row respectively. To facilitate a fair comparison, the clusters presented adhere to the lowest $\mathcal{M}_{2}$ criterion already imposed. Shown in Figure ~\ref{fig2}b and in a similar fashion to the cubic clusters we find the progression closely follows the evolution seen in non-magnetic spherical colloidal clustering from experiments\cite{Cho:2005cp}. Again, this identification is relative to the center of mass for the spheres. The insights from the previous section regarding the dominance of confinement over magnetic forces are valid once again. Turning our attention to the lower row with dipoles, we can already visually identify configurations with significantly more ordering of the magnetic moments than those observed for cube clusters. Ring formation has reasserted itself, moreover, we see the appearance of distinct layers in the configuration of the dipoles. One can argue that these begin to manifest from $n\geqslant{4}$, starting with two layers of anti-parallel pairs. Due to the prevalent return of flux closure in these clusters, we expect the remanent magnetisation to be less in comparison to the equivalent cube clusters.
\\
\\
 %===============
% Cluster Comparison
%===============
\noindent \textbf{Cluster Comparison}
\\

\begin{figure*}
	\centering
	\includegraphics[width=0.8\textwidth]{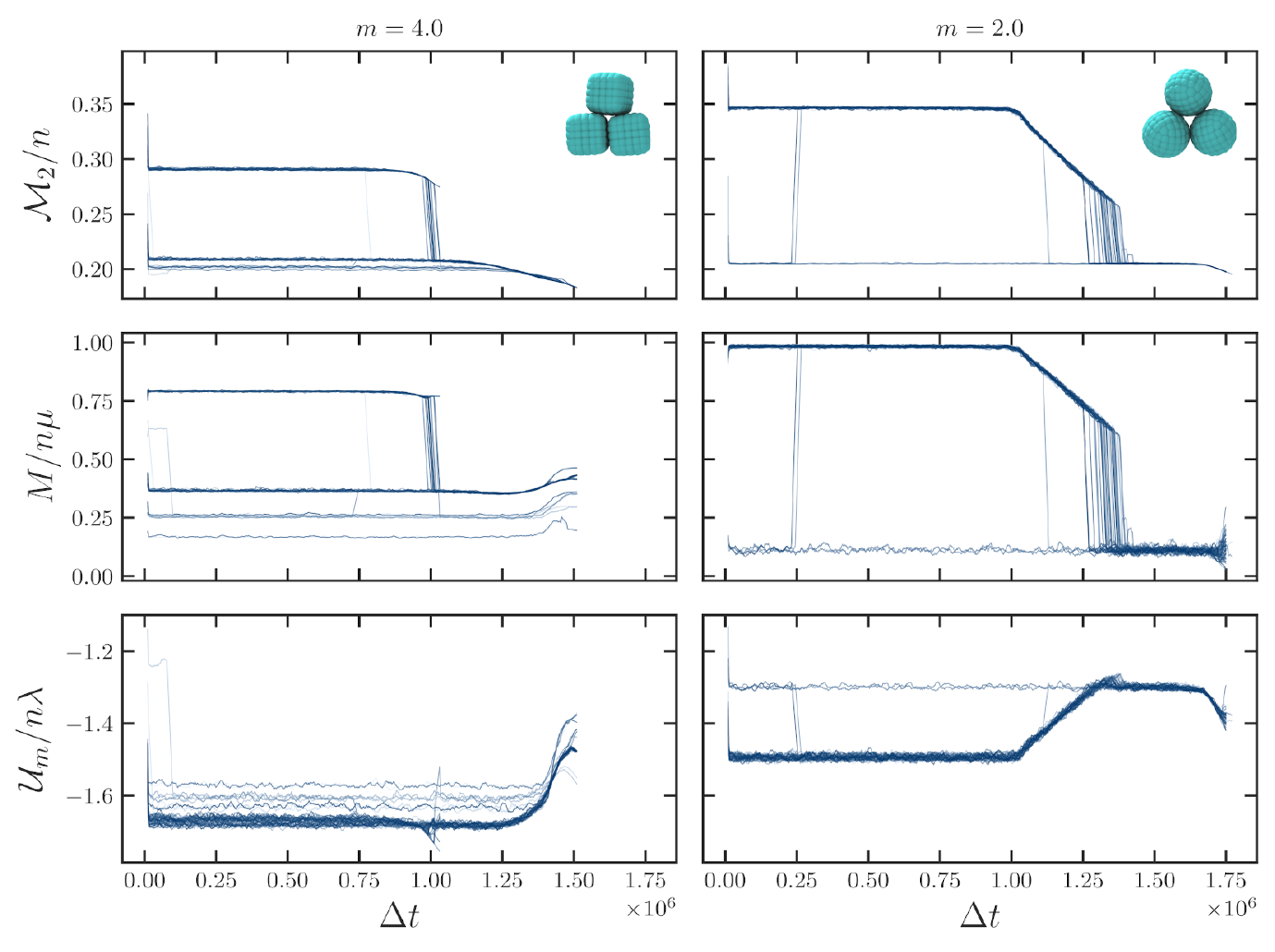}
	\caption{
	{\bf Cluster Property Comparison.}
	Plots of the measures used to describe and monitor the evolution of cluster formation. The data shown is for a cluster size of $n=3$. Equivalent plots for each of the other cluster sizes can be found in the Supporting Information (Figures~S1-S8). The grid of plots is arranged as follows, each column displays the data for each particle type, cubes ($m=4$)  and spheres ($m=2$) on the left and right respectively. In the upper row of plots we have the second moment of the mass distribution, (the cluster selection criterion), followed below by the total dipole moment of a cluster, and ending with the magnetic interaction energy across the whole cluster. Each plot shows the evolution of the respective quantity over the course of a simulation, the evolution is plotted in units of the simulation time-step $\Delta t$. Each quantity is normalised in the manner indicated to facilitate comparisons not only between particle types but also cluster sizes, where as a reminder, $\mu=|\boldsymbol{\mu}|$ is the particle magnetic moment, $n$ is the cluster size, and $\lambda$ is the magnetic coupling parameter (see Methods). Fifty replica compression runs were performed for each type of cluster for the given particle size. To aid further with readability, the evolution of each replica was smoothed by calculating the moving average over 200 measures. 
	}
    \label{fig3}
\end{figure*}

\noindent In the preceding analysis we selected a single cluster from the set of replicas for each value of $n$ according to the minimal $\mathcal{M}_{2}$ criterion. In contrast to this specificity, we will now address quantitatively the variability across all replicas for each cluster size and make comment on the reproducibility of the structures discussed thus far. Figure~\ref{fig3} shows the three quantities used for the analysis and comparison of cube (left column) and sphere (right column) clusters. Namely, 

\begin{align}
\label{eqn:2ndMoment}
\mathcal{M}_{2}&=\sum^{n}_{i=1}(\boldsymbol{r}_{cm}-\mathbf{r}_{i})^{2},
\end{align}
\begin{align}
\label{eqn:mag}
%M=|\bf{M}|&=|\sum_{i}^{n}\boldsymbol{\mu}_{i}|,\\
M&=|\sum_{i}^{n}\boldsymbol{\mu}_{i}|,\\
\label{eqn:magerg}
\mathcal{U}_{m}&=\sum_{i,j}^{n}U_{m}(r_{ij},\boldsymbol{\mu}_{i},\boldsymbol{\mu}_{j}).
\end{align}
$\mathcal{M}_{2}$ is the second moment of the mass distribution, where $\boldsymbol{r}_{cm}$ represents the center of mass of the cluster and $\boldsymbol{r}_{i}$ the location of each individual surface site, which allow for the geometry and orientation of the particles to be implicitly accounted for. $M$ denotes the scalar magnetisation (or total dipole moment) of the cluster. $\mathcal{U}_{m}$ is the total magnetic interaction energy where $U_{m}$ is the  dipole interaction between two particles $i$ and $j$ as defined in the Methods section. These observables are plotted as a function of the time evolution of the simulations, {\it i.e.} the progression as the droplet evaporates, expressed in terms of the number of time-steps $\Delta t$. Each observable is normalised in a manner that allows the data for different cluster sizes and particle types to be viewed on an equal footing. We present here the evolution for $n=3$, a cluster type that we will explore the assembly of later in this work. Equivalent datasets for all other cluster sizes investigated are presented in the Supporting Information (Figures~S1-S8). 

% ==== Figure 3 ====
 
To begin let us consider each particle type separately. For cubic particles (Figure~\ref{fig3} column 1), $\mathcal{M}_{2}$ (row 1) for each replica converges to the same value, indicating that the same structural arrangement of particles is being reproduced in a regular, repeatable fashion. $\mathcal{M}_{2}$ provides a measure of the distribution of the particles in the cluster and thus a measure of how the particles are arranged in space. Following the evolution of $M$ (row 2) we observe a lack of convergence over the course of confinement. $M$ describes the magnitude of the cluster magnetic moment, an indication of the remanent magnetisation {\it i.e.} the propensity of a cluster to maintain magnetic character. One can conclude then that although replicas readily form equivalent structural arrangements, the spread in the remanent magnetisation of the resultant clusters suggests the dipoles within a cluster must be oriented differently. This is further corroborated by considering $\mathcal{U}_{m}$ (row 3), the total magnetic interaction energy, where we again note a deviation in the final values. This suggests that either the distance between, or orientation of, the dipoles is varying within the clusters. However, we know that the cluster symmetry is consistent from the evolution of $\mathcal{M}_{2}$ implying that it is strictly the dipole orientations that are inconsistent from cluster to cluster. Turning our attention to spheres (Figure~\ref{fig3} column 2) one notices immediately the tendency for each replica to converge to broadly similar values for all three measures. The fluctuations in the closing stages of the evolution in $M$ and $\mathcal{U}_{m}$, appearing from a clearly previously well-defined pathway, can be attributed to the lower structural rigidity of the sphere trimer. The structure can be deformed more easily by the evaporating droplet than its cube counterpart, which partially stabilises itself due to steric hindrance. Prior to this deviation, the values between replicas are broadly self-consistent.

Comparing between the particle types we note the similarity in the values of $\mathcal{M}_{2}$, suggesting the equivalency in the structural arrangements for both cluster types, emphasised by the inset snapshots of Figure~\ref{fig3}. For the two magnetic parameters, we can see a clear-cut spread in the values for cubes and the pathways to arrive there, this is not the case for spheres where a much clearer consistency is found. This allows us to conclude that the spherical particle clusters offer the best opportunity to not only reliably and reproducibly attain a consistent cluster geometry but also reliably reproduce equivalent magnetic configurations and characteristics. For further confirmation and evidence of these assertions, the reader is encouraged to study the equivalent plots for $n=2,(4-10)$ that appear in the Supporting Information (Figures~S1-S8), where similar behaviour is seen across clusters with different values of $n$. 
If one looks at the pathways taken by the respective particle types during confinement, cubes proceed \textit{via} multiple possible trajectories due to the complex free energy landscape generated by the competition between steric and magnetic interactions. In contrast, spheres proceed by one clearly defined pathway characterised by two branches, visible in each of the observables: the upper branch corresponds to spheres in a chain configuration, the chain then deforms, buckles and collapses to the lower branch which indicates flux closure and the formation of a ring. The closure of the ring occurs at different points in time for each replica as determined by the confinement and the random Brownian fluctuations. During compression the vast majority of dipolar rearrangement occurs concurrent to the cluster formation. Once a particular structural arrangement is formed during cluster formation this has a corresponding dipolar arrangement as determined by the trajectory of the simulation prior to the "collapse" into the cluster. Consequently, one can say that the cluster formation and dipole rearrangement take place on the same timescale. This follows on from the fact that the dipoles within the particles are fixed relative to their geometry. The increase in the potential energy at long times seems to suggest that there is still some rearrangement of dipoles after the clusters have formed, however this can be attributed to perturbations of inter-particle separation as a result of compression, which leads to the fluctuations seen in the observables. The particles are being forced closer together, causing the increase in energy prior to the simulation end. This is not due to further drastic alterations in orientation of dipoles relative to the cluster geometry in which they are present.

% ==== Figure 4 ====
\begin{figure}
	\raggedright
	\includegraphics[width=0.45\textwidth]{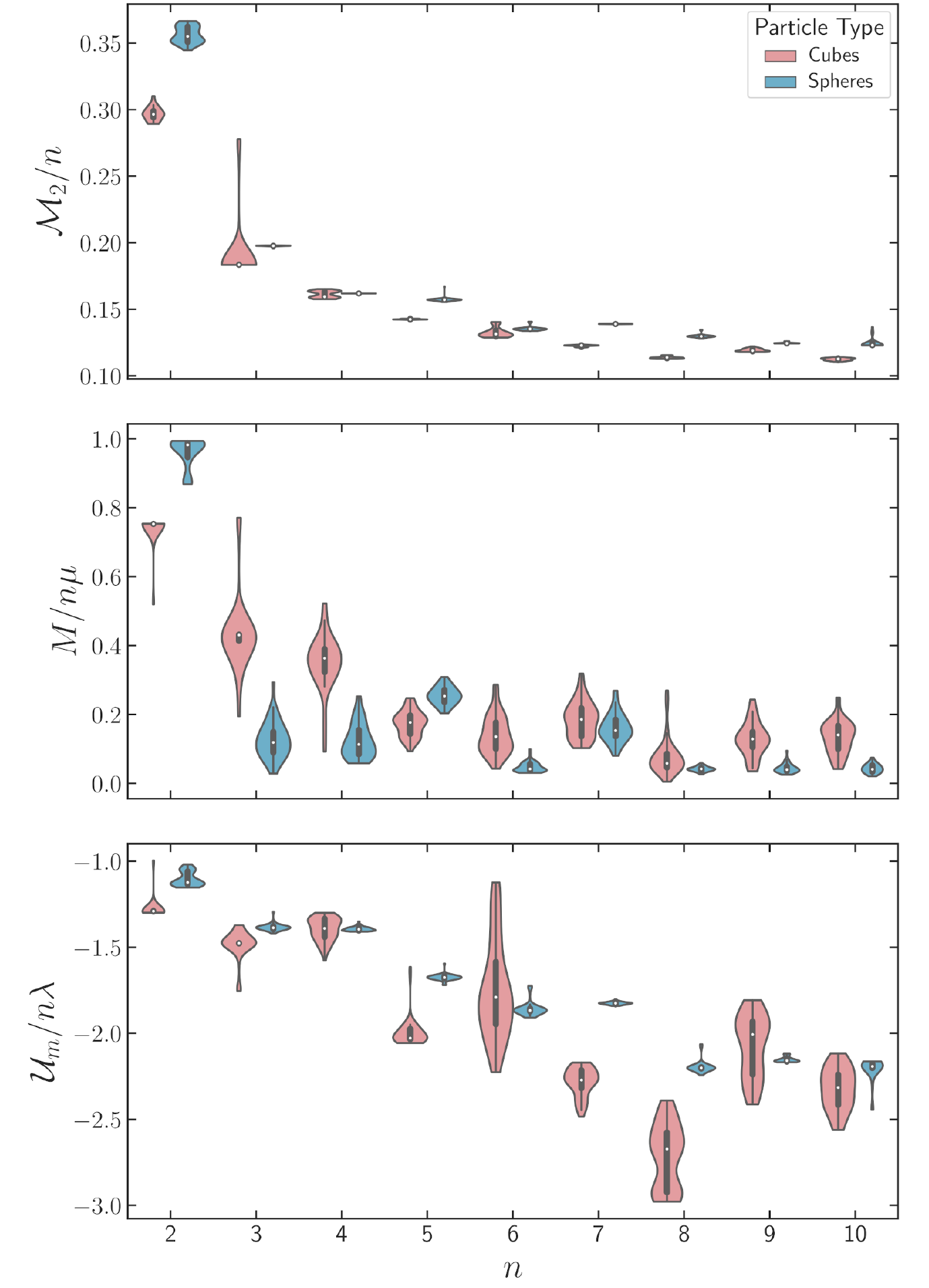}
	\caption{
	{\bf Cluster Property Distributions.}
	In three violin plots we summarise the observable of interest as a function of cluster size $n$, for every replica at the end of the evaporation procedure. In the upper plot we present the second moment of the mass distribution $\mathcal{M}_{2}$, in the middle plot we look at the cluster magnetisation magnitude $M$, and in the lower plot we look at the total dipole interaction energy across the cluster $\mathcal{U}_{m}$. We maintain the same normalisation strategy as discussed for Figure~\ref{fig3}. Distributions for sphere particle clusters are shown in blue, while cube cluster are shown in red. The distributions drawn take into account only the available data and thus truncate at its limits. A boxplot is drawn at the centre of each distribution where the white circle denotes the median, the black bar denotes the interquartile range and the black line denotes the maximum and minimum extent neglecting outliers. Viewing the data in this manner confirms that while all clusters of cubes and spheres show reproducible structural configurations, only clusters of magnetic spheres show reproducible magnetic configurations. }
    \label{fig4}
\end{figure}

We can go one step further in our analysis and facilitate a more quantitative comparison of the resultant cluster geometry. For all replicas of a given cluster size we collated the terminal values of each of the three observables. We summarised this data in the form of a violin plot appearing in Figure~\ref{fig4}, in which individual distributions of $\mathcal{M}_{2}$, $M$, and $\mathcal{U}_{m}$ are visualised for each cluster size $n$, where data for spheres appear in blue and cubes in red. Each violin shows the probability density in the horizontal plane and the quantity under consideration varies in the vertical plane. In terms of the second moment of the mass distribution, we see that the structural similarity between clusters of spheres and cubes is very strong, and the values of $\mathcal{M}_{2}$ are in close proximity for a given cluster size. Furthermore, we note that the spread of the values in either case is predominately very narrow, highlighting the reproducibility of the structural arrangement of the clusters in space. In general, the decrease in $\mathcal{M}_{2}$ with increasing cluster size indicating an increase in the spherical symmetry of the clusters. Considering next the magnetisation (total dipole moment) of the clusters in the middle plot, the most notable difference to the previous quantity is that there are now much broader distributions in the values for each cluster size and particle type, this width does decrease for the spherical case as the cluster size increases. Moreover, the size of the spread is in general less for clusters of spheres. These observations are indicative of the fact that we have more variation in the dipolar configurations achieved upon compression particularly so for the cubic particles. In the spherical case we see a propensity for the clusters to do a better job of closing the magnetic flux within the cluster, minimising it close to zero as cluster size increases. This highlights the magnetic frustration felt by the cubic clusters on compression due the the steric hindrance generated by the cubic geometry. 
The magnetic energy offers complementary insights into the magnetic configurations. In this case, we notice that the energies of the sphere clusters are distributed in a much narrower fashion in comparison to the cube counterparts. The energy per particle is seen to broadly decrease with growing cluster size, discontinuities in this trend are likely due to the frustrations induced by an additional particle being difficult to incorporate in the previous structure type. Care should be taken when comparing cluster energies between particle types due to the variation in particle dimension that result from the fixed volume of the particles. It is not out of the question that although a given sphere cluster is both structurally and magnetically favourable the corresponding cube cluster, although structurally equivalent but magnetically frustrated, could be lower in energy simply do to the fact that the dipoles are slightly closer together. If we consider the magnetisation and magnetic energy simultaneously we believe we can offer an explanation for the spread in the magnetisation observed for both particle variants. In the spherical case, the tight spread of cluster energies implies the dipoles are likely to be broadly in the same orientation within a given cluster, the modest variation in the magnetisation is thus likely due to the fluctuations of the dipoles around these given directions. Fluctuations are possible due to the sphere's ability in the simulation to rotate freely even while bound in the cluster. In experiments however, even if rotations are hindered by van der Waals forces between adjacent particles, we would expect a similar distribution in the magnetisation due to thermal fluctuations acting during compression prior to irreversible aggregation.
We do not expect these minor differences between clusters to inhibit the subsequent hierarchical assembly pathways.
Contrastingly, for cubic clusters, the variation in cluster energy is predominately due to dipoles becoming fixed in different orientations within the structure. Once in a cluster, the rotational freedom for the cubes is constrained by the presence of the other particles in the arrangement, consequently fluctuations of the dipole around the average rotation are lessened in comparison to spheres. This observation suggests that the variations in magnetisation for cubic clusters are due to manifestly different dipole orientations and thus configurations of cubes within a cluster. This further cements the previous qualitative observations that clusters of spheres are far better at reproducing not only the structural arrangement in space but also the magnetic arrangement. Our cubic systems can only reproduce the former on a consistent basis. We therefore suggest that the spherical variant is the most viable candidate for producing a colloidal hierarchy of magnetic building blocks. In simple terms this mean that we should theoretically be able to produce clusters of spheres with consistent shape and magnetic configuration to be used for hierarchical assembly. 

% ==== Figure 5 ====
\begin{figure*}
	\centering
	\includegraphics[scale=0.95]{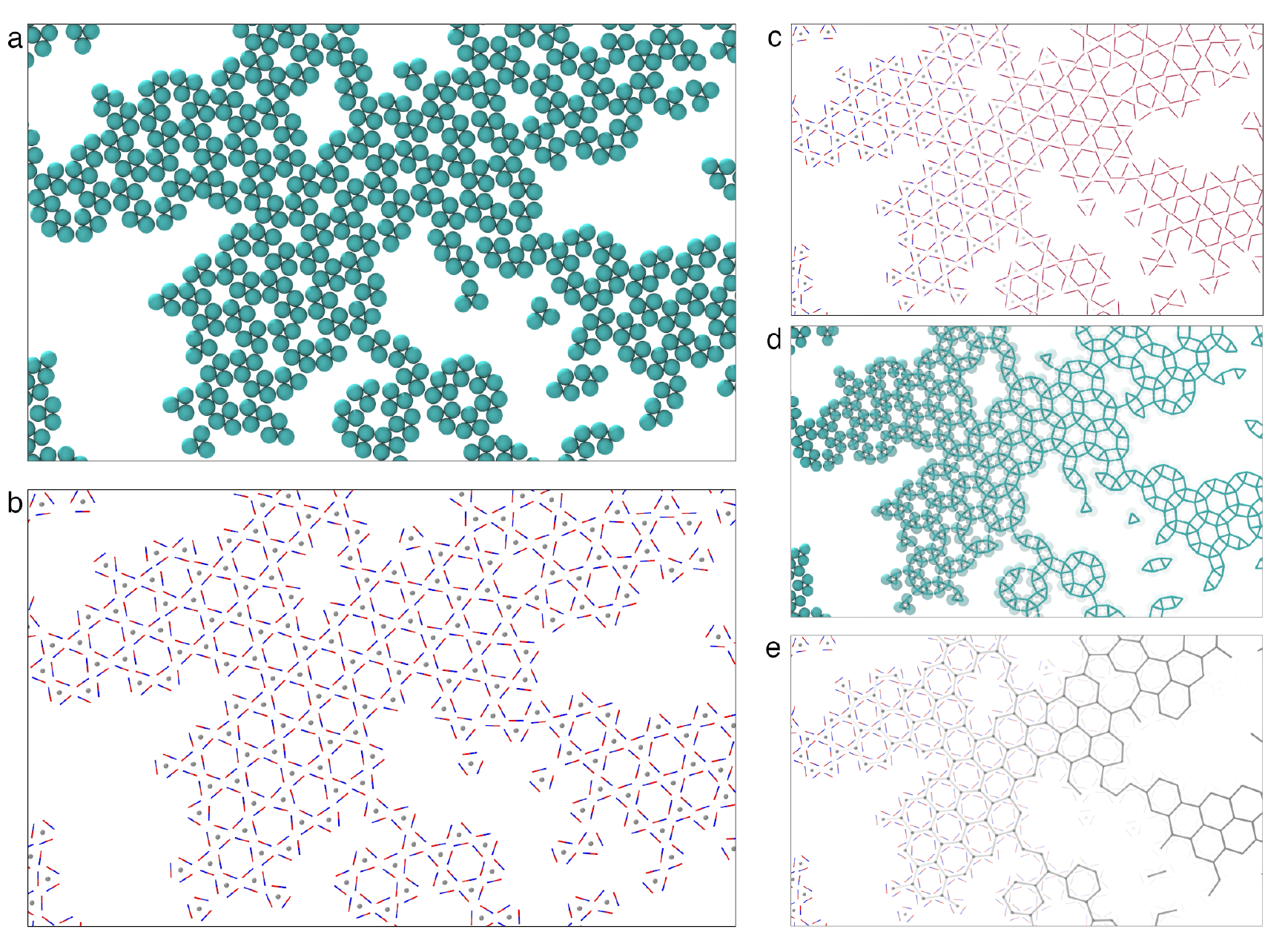}
	\caption{
	{\bf Cluster Aggregation}
	A single snapshot from the monolayer simulation at a concentration of $\varphi_{A}=0.4$ for clockwise trimers. The field of view within the simulation has been reduced to allow more detail to be seen, a complete field of view of the simulation can be found in Figure S9 of the Supporting Information. Each image (a-e) is of the same region within the monolayer. (a) Main structural arrangement of the clusters. (b) We peer inside the clusters here, highlighting the arrangement of the dipoles (red-blue bar) within, the center of mass of each cluster is indicated by the grey sphere. In (c-e) we showcase the different Archimedean lattice structures co-existing within the monolayer. In these images we transition from the relevant snapshot image (left) to a simplified visualisation of the lattice (right) to highlight the repeating pattern. (c) A kagome lattice formed by the arrangement of the dipoles in the monolayer structure. (d) A bounce lattice formed across the monolayer by the individual particles constituting each cluster. (e) A honeycomb lattice in the monolayer formed by considering the centre of mass of each cluster. Corresponding images for the anticlockwise and racemic systems can be found in Figure S10 $\&$ S11 of the supporting information respectively.}
    \label{fig5}
\end{figure*}

 %===============
% Hierarchical asssembly
%===============
\noindent \textbf{Hierarchical assembly}

To confirm the validity of the previous observations, we have run simulations to test the hierarchical assembly capabilities of magnetic trimers, clusters formed by three magnetic spheres. In the interest of simplicity, the trimers were considered idealised versions of that appearing in Figure~\ref{fig2}b. Namely, the center of mass of each sphere was placed at the vertex of an equilateral triangle defined by an edge length equal to the sphere diameter. The dipoles were oriented perpendicular to the displacement vector for each sphere, relative to an origin at the triangle centroid. The trimers were confined to a strictly two dimensional monolayer, where cluster rotations were only permitted in plane. Simulations were conducted on a bulk system where the number of clusters was $N_{c}=1000$. Periodic boundary conditions were employed to mimic the bulk of a monolayer. The system was initialised by placing clusters at random positions and orientations at an area fraction of $\varphi_{A}=0.4$. Furthermore, due to the imposed two-dimensional system geometry we arrived at a situation where clusters can be considered as magnetic enantiomers of one another. To account for this effect, three systems were propagated to see the effects on their assembly. Two scenarios with systems of clusters of one type were used, namely, where the dipole configuration circulated in a clockwise and anticlockwise direction respectively. We adopt here a naming convention that follows the blue end of the dipole visualisation in the simulation snapshots. The third scenario considered was a racemic mixture of both cluster varieties. Further details on the simulation method used to explore the cluster aggregation can be found in the Methods section. 
Analysing the trajectories taken by the three systems we could quickly identify the clockwise and anticlockwise systems evolved in an equivalent fashion, whereas pattern formation in the racemic mixture was frustrated due to the different enantiomers being present. Nevertheless, enantiopure crystallites are beginning to emerge as islands within the bulk (see Figure~S11). 
Experimentally it is not unreasonable to anticipate phase separation of enantiomers in 2D samples given enough equilibration time. Furthermore, non-uniform magnetic fields could be used to separate enantiomers or to prepare enantiopure samples by enforcing a certain orientation of each trimer. One should note however that, at least for the trimers, chirality is lost in 3D.
Taking the clockwise variant as an example of an enantiopure system, the results of the cluster aggregation are shown in Figure~\ref{fig5}, where we have a cropped view of the simulation cell, a full view can be found in Figure S9 of the Supporting Information. For Figure~\ref{fig5}a-b we see the positioning and dipolar arrangement of clusters in the aggregated structure respectively. In Figure~\ref{fig5}c-e we compartmentalise the repeating patterns found in the aggregated monolayer to highlight a number of Archimedean lattices that manifest in different aspects of the structure. These images take a gradient from the respective structural snapshot and morph gradually into a simple rendering of the lattice we wish to highlight. It is clear from Figure~\ref{fig5}a that we have the formation of a hierarchical well-ordered lattice structure, in which point defects and dislocations are still evident. Point defects manifest as holes in the lattice where one or two trimer units is missing. Dislocations occur between ordered crystallites and result in the formation of alternating five and seven membered rings in contrast to the more energetically advantageous six, this is most clearly seen in the upper right hand portion of Figure~\ref{fig5}c. It should be noted that this structure formed spontaneously under the simulation condition, with no use of more sophisticated simulation techniques to optimise the structure. The characteristic motif within the structure is evidently the interlinking six-membered rings. Turning to Figure~\ref{fig5}b we visualise the dipoles within each cluster. The center of mass for each cluster is indicated by a silver sphere to act as a reference and to aid with the comparison to the other visualisations. One can note that the dipolar configuration is characterised by archetypal ring formation, albeit with a hexagonal flavour. Considering now the ordering of the clusters within the monolayer we can make a number of identification of how the aggregate repeats in space. 
The pattern arsing from dipole alignment can be characterized as a staggered kagome lattice, as shown in Figure~\ref{fig5}c. This lattice is not a true kagome lattice, as the vertices of the triangles formed by connecting the particle dipoles overlap, disturbing the exact trihexagonal tiling present in a true kagome lattice. In Figure~\ref{fig5}d, by considering the constituent particles of each cluster as lattice points, we find the particles arrange themselves into a so-called bounce lattice. Finally, if we treat the center of mass of each cluster as a lattice point, we find a honeycomb lattice as shown in Figure~\ref{fig5}e. 
Having broken down the repeating structure of the monolayer into its constituent parts, it is clear to see the complex ordering one can obtain in both the topological and magnetic characteristic of the monolayer. The repeating lattice patterns present in the monolayer are well understood and quantified, however the bounce lattice, in particular, has not yet been seen or predicted in colloidal systems including in experimental and theoretical works on patchy colloids\cite{Doppelbauer:2010iv,Chen:2012fm,Zhang:2015fe,Chen:2011be,Elacqua:2017jp,Noya:2017cd,Bianchi:2017bt,Li:2020hc}, which are the most closely related systems available as of yet.
The observed structures are strikingly different and of greater complexity compared to those obtained from the assembly of the simple dipolar spheres, the "monomers" of our hierarchical structures. These are in fact known to form ring and chain structures at low concentrations\cite{Ivanov:2015ff}, branched structures at intermediate concentrations\cite{Kantorovich:2015cw,Rovigatti:2013ku}, and close-packed structures at higher concentrations\cite{Spiteri:2017ea,Messina:2015fh}.
 Our structures can therefore only be accessed using hierarchical assembly: constituent magnetic particles pre-assembled into a larger unit, a building block, the structure and magnetic configuration of which directly influence the subsequent level of assembly where the building blocks organize to form the ordered monolayer. In the case of trimers, we have clearly shown proof of concept for such a protocol with this kind of spherical magnetic particle. This route offers the possibility of engineering hierarchical colloidal materials that are magnetically reactive.

%=============
%=============
% Conclusions
%=============
%=============
\section*{Conclusions}

In this work, we have introduced \textit{via} computer simulation a viable way to prepare colloidal magnetic building blocks by confining magnetic cubes and spheres into small clusters. While the lower symmetry of the magnetic cubes frustrates the magnetic arrangement during confinement, clusters made of magnetic spheres show exquisitely reproducible magnetic configurations for clusters of up to ten particles. We have shown that magnetic sphere trimers (clusters made of three magnetic spheres) readily assemble into ordered monolayers in which three of the eleven Archimedean lattice symmetries can be identified. We anticipate the experimental analogs of our clusters to be stable in dispersion due to strong van der Waals forces arising upon particle contact, comparably to other already available experimental systems \cite{Manoharan:2003hb,Yi:2004vo,Cho:2005cp,Peng:2013bb,PazosPerez:2012bu}. The method presented in this work has therefore the potential to open alternative avenues for colloidal self-assembly using building blocks that can be prepared in bulk and interact with highly specific interactions without the need of additional costly chemical functionalisations.

%=========
%=========
% Methods
%=========
%=========
\section*{Methods}

%=========
% Computer Simulation
%=========
\noindent \textbf{Computer Simulation}
\\
\\
% ==== Model ====
The particles in this work were constructed from sub-units of spheres using a real and virtual particle scheme to encapsulate rigid body motion. A real site is placed at a particle's center of mass, relative to which virtual particles are positioned, building up the particle surface. The details of this scheme for particle construction are discussed in detail in Ref.~\citenum{Donaldson:2017ns}. In contrast to the previous work the positioning and sizing of the sites comprising the particle surface has evolved. The surface is constructed from overlapping spheres of equal diameter, positioned equidistantly from each other on a lattice lying at the boundary defined by the following equation describing the geometry of the superball surface,
\begin{equation}\label{eqn:superball}
\left(\frac{2x}{h}\right)^{m} + \left(\frac{2y}{h}\right)^{m} + \left(\frac{2z}{h}\right)^{m}= 1,
\end{equation}
where $h$ is the height of the particle and $m$ is the shape parameter that sets the roundness of the particle edges and vertices\cite{Batten:2010dd}. The diameter of the surface sites was set by the number of sites used relative to the lattice spacing. The surface particles were placed on the boundary according to the routine outlined for the surface charges appearing in Ref. \citenum{Rosenberg:2020bs}. At the coordinates of each surface site the normal to the surface was calculated according to,

\begin{equation}\label{eqn:normal}
\begin{aligned}
&F=\left(\frac{2x}{h}\right)^{m} + \left(\frac{2y}{h}\right)^{m} + \left(\frac{2z}{h}\right)^{m}-1;\\ \\
&\qquad
\mathbf{\hat{n}}(x,y,x)=\frac{\boldsymbol{\nabla} F}{|\boldsymbol{\nabla} F|}.
\end{aligned}
\end{equation}

The particle was then shifted by $\tfrac{h}{2}$ in the direction of $-\mathbf{\hat{n}}$. In this manner, the edges of surface sites lie on the boundary defined in Equation~\ref{eqn:superball}. The number of surface sites used is equal to 150, {\it i.e.} 25 per face in the case of a cube particle. This number was determined based upon a trade off between efficacy and accuracy.

We have studied superball particles with $m=2$ (spheres) and $m=4$ (cubes) exclusively. The shape of the cubic magnetic particles is based on those appearing in Ref.~\citenum{Rossi:2018ef} that are composed of hematite. The magnetic character of hematite particles can be suitably approximated by a dipole placed in the centre of the superball. Similarly we use the dipole moment orientation reported therein, namely a 12$^{\circ}$ tilt from the space diagonal towards the cube face. The dipole orientation relative to the sphere geometry is irrelevant due to the symmetry present. In the experimental system hematite superballs with $m=4$ had a height of $h=(L+2t)=1335\text{nm}$ where $L=1135\text{nm}$ denotes the height of the magnetic core and $t=100\text{nm}$ was the thickness of a silica shell. At this point it is useful to define a number of pertinent reduced units used during simulations. Namely, temperature as $T^{*}=kT/\epsilon$, magnetic moment $({\mu^{*}})^{2}=\mu_{0}\mu^{2}/4\pi h^{3}\epsilon$, energy $U^{*}=U/\epsilon$ and displacement $r^{*}=r/h$. Where the following identifications are made: $k$ the Boltzmann constant, $\epsilon$ the energy parameter, and $\mu_{0}$ vacuum permittivity. In these simulation units the particle height becomes $h^{*}=1$. This results in a superball volume of $\nu_{sb}^{*}(m=4.0)=0.810248$. It follows that for $\nu^{*}_{sb}(m=2)=0.810248$ we require $h^{*}=1.156662$. This scaling correlates with the behaviour in experimental systems as the magnitude of a particle's magnetic moment scales with the volume of the particle $|\boldsymbol{\mu}|\propto{\nu}$. We keep $\nu^{*}_{sb}$ constant when moving from cubes to spheres, a restriction that is compensated for by an increase in the sphere diameter. In other words, we created a spherical analogue to the established cubic particles. Illustrations of the particle models for spheres and cubes are found in Figure~\ref{fig1}b and \ref{fig1}c respectively.

We can link the simulation and experimental realms by characterising the system using the magnetic coupling parameter,
\begin{equation}
\begin{aligned}
& \lambda=\frac{\mathcal{F}\mu_{0}\mu_{p}^{2}}{8\pi(L+2t)^{3}kT}=\frac{\mathcal{F}(\mu^{*}){^2}}{2T^{*}};\qquad  \\  \\ & \text{where }\\ \\& \mathcal{F}=\frac{5+\cos\left(2\left[\theta+\cos^{-1}\left(\frac{\sqrt{6}}{3}\right)\right]\right)}{4}\approx1.2303\text{ for }\theta=12^{\circ},\\
\end{aligned}
\end{equation}
relating the magnetic and thermal energy\cite{Rossi:2018ef}. The quantity $\mathcal{F}$ is a structural pre-factor relating to the dipole tilt angle $\theta$ and the two particle ground state. An experimental value of $\lambda$ was calculated for the cubic particles discussed, with $T=100^{\circ}$C (temperature of the system during droplet evaporation) and $\mu_{p}=2.8\times10^{-15}Am^{-2}$ (for hematite), resulting in $\lambda=39.3435$. By choosing $T^{*}=1$ in simulations, the corresponding magnetic moment was calculated as ${\mu^{*}}=7.99735\sim8$ and used for both particle types. The short range interaction between particles was treated as the sum of repulsive contributions between each spherical sub-unit, characterised by the Weeks-Chandler-Anderson potential,
\begin{equation}\label{eqn:wca}
	U_{s}(r)=
		\begin{cases}
  			4\epsilon\left[(\frac{\sigma}{r-r_{\text{off}}})^{12}-(\frac{\sigma}{r-r_{\text{off}}})^{6}\right]+ \epsilon, &   r <r_{c}+r_{\text{off}}\\
			 \hspace{2cm}0, &r\geqslant r_{c}+r_{\text{off}}
  		\end{cases},
\end{equation}
where $r$ is the displacement between surface sites on opposing particles and $\sigma$ denotes the surface site diameter and energy parameter $\epsilon$ defines the energy scale. The cut-off radius $r_{c}$, at which the interaction potential becomes zero, is defined to be $r_{c}=2^{1/6}\sigma$. An offset radius $r_{\text{off}}$ was employed to tune the location where the potential falls to zero. In order to steepen the potential, making it less soft, we used $\sigma$ and $r_{\text{off}}$ in tandem to achieve this. Namely, we actually mirror a hard particle diameter of $\sigma$ by setting $\sigma=R$ and $r_{\text{off}}=R$, where R is the virtual site radii. This produces a steeper more hardcore potential that still falls to zero beyond $\sigma$. The magnetic interaction is approximated using the dipole potential,
\begin{equation}\label{eqn:dip}
	U_{m}(\mathbf{r},\boldsymbol{\mu}_{1},\boldsymbol{\mu}_{2})= \frac{\mu_{0}}{4\pi}\left[\frac{\left(\boldsymbol{\mu}_{1}\cdot\boldsymbol{\mu}_{2}\right)}{r^3}-\frac{3\left(\boldsymbol{\mu}_{1}\cdot\mathbf{r}		\right)\left(\boldsymbol{\mu}_{2}\cdot\mathbf{r}\right)}{r^5}\right],
\end{equation}
where $\mathbf{r}$ denotes the vector between dipoles $\boldsymbol{\mu}_{1}$ and $\boldsymbol{\mu}_{2}$, with a magnitude of $r=|\mathbf{r}|$.
\\
\\
% ==== Protocol ====

\begin{figure}%[t]
	\centering
	\includegraphics[width=0.45\textwidth]{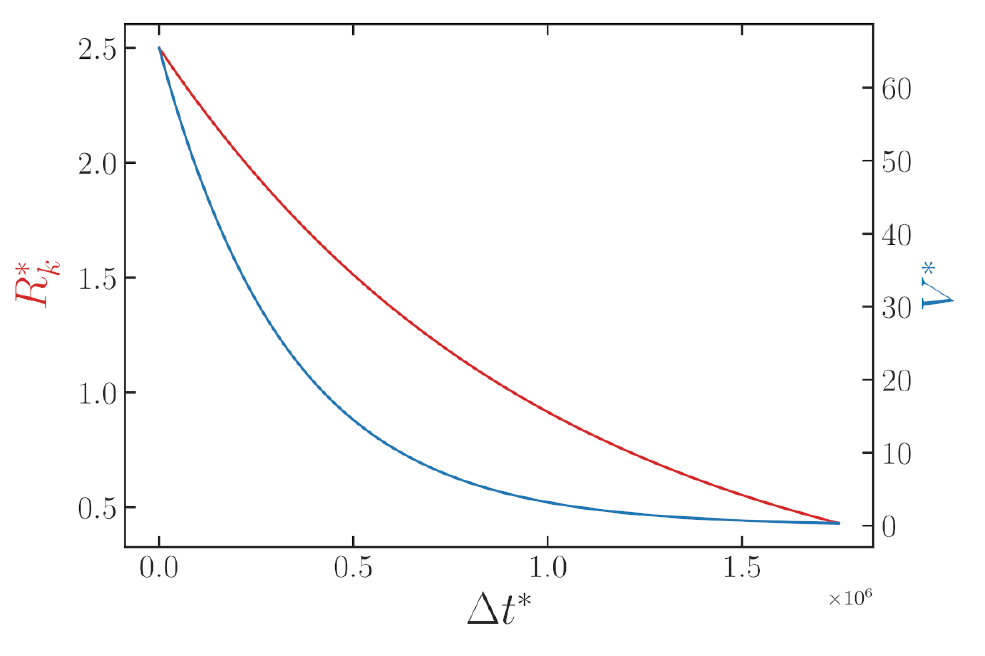}
	\caption{
	{\bf Droplet Evaporation}
	Visualisation of the droplet evaporation scheme used in simulation for a cluster size of $n=3$. The droplet radius $R_{k}$ is systematically decreased over the course of the simulation according the the curve appearing in red. The corresponding reduction in the droplet volume is shown in blue. The curves are both plotted as a function of the time-step $\Delta t^{*}$}.
    \label{fig6}
\end{figure}

\noindent \textbf{Droplet Evaporation.} Simulations were conducted on isolated clusters of particles ranging in size from $n=(2-10)$, for both $m=2,4$. Individual runs were initialised by randomly distributing in both position and orientation $n$ particles confined to the inside of a sphere of radius $R_{i}^{*}=\tfrac{(n+2)}{2}$, within a three dimensional non-periodic simulation box. The sphere is present to imitate the evaporating droplet from the experimental systems alluded to in the main text. The surface sites of particles also interacted with the confining sphere \textit{via} the potential in Equation~\ref{eqn:wca}, where in this case $r$ is the displacement between site centres and the droplet surface. The initial sphere radius $R_{i}^{*}$ was chosen sufficiently large to not preferentially bias the system into any particular area of the free energy landscape. The system was propagated according to Langevin molecular dynamics, the use of which in this context is discussed in detail in previous studies \cite{Donaldson:2015ns,Donaldson:2017ns}. Due to the non-periodicity of the system the dipolar interaction was calculated using direct summation. As noted earlier all simulations were conducted at $T^{*}=1$ and with particle magnetic moments of $\mu^{*}\sim8$. The time step used was $\Delta t^{*}=0.001$. During the course of a single cluster simulation the confining sphere was reduced in size according to the following equation, 
\begin{equation}\label{eqn:drop}
R^{*}_{k}=R^{*}_{i}(0.99)^{k},
\end{equation}
where $R_{k}$ is the radius after $k$ iterations. 
In Figure~\ref{fig6}, we plot the variation of $R_{k}$ (red) over the course of a simulation for a cluster size of $n=3$ and as a function of $\Delta t^{*}$, alongside we plot the corresponding droplet volume (blue) given by $V^{*}=\tfrac{4\pi {R_{k}^{*}}^{3}}{3}$.
Setting a rate constant of 0.99 ensures the particles contained are confined gradually, and able to stay in a quasi-equilibrium state. This scheme approximates the gradual evaporation of the water from the droplets in experiment. One can view this as a simulated annealing protocol, which instead of acting on temperature acts on the sphere size. Using this scheme meant that the reduction in droplet size at each iteration was reduced as the simulation progressed. By maintaining the iteration length, the confinement was applied more slowly as the system increased in density and thus harder for rearrangement to occur. This allows the free energy landscape to be properly explored especially when replica simulations are used. In this case 50 replicas were performed for each value of $m$ and $n$. After each reduction in droplet size or k\textsuperscript{th} iteration the system was propagated for $2.0\times10^{4}\;\Delta t^{*}$ to allow for equilibration. The evolution of droplet evaporation was observed and recorded: observables (energy {\it etc.}) every $1.0\times10^{2}\;\Delta t^{*}$ and particle configurations once immediately prior to the next confinement iteration. Simulations were stopped when the force on the confining sphere was seen to diverge, {\it i.e.} the point at which the particles begin to penetrate the confinement. A schematic of the procedure using real simulation data is shown in Figure~\ref{fig1}a. It should be stressed that the compression procedure was the same for both particle types, meaning the relative difference in the magnetic structure and particle arrangement are comparable. 

From the 50 replicas given for each $m$ and $n$ the one achieving the lowest value of the second moment of the mass distribution (Equation~\ref{eqn:2ndMoment}) was selected for visualisation. In previous studies this was reported as a effective parameter with which to differentiate clusters \cite{Cho:2005cp,Yi:2004vo,Manoharan:2003hb}. Simulations in this study were performed using ESPResSo 3.3.0 \cite{Arnold:2013esp}. Similar simulation schemes to this \textit{i.e.}  at constant volume in the NVT ensemble have been shown to achieve indistinguishable results to those conducted using the NPT ensemble\cite{Wang2018:nc}.

% ==== Protocol ====
\noindent \textbf{Cluster Aggregation.} For the simulations of spherical particle trimers we abandoned the use of the composite sphere model discussed above, and reverted to a simple dipolar soft sphere implementation characterised by the potentials in Equation~\ref{eqn:wca} and Equation~\ref{eqn:dip}. This choice was made to improve the efficacy of the simulations and absence of the need to compare to the cubic case. Moreover, the magnetic moment of the particles was reduced to $\mu^{*}=2.5$, whilst the temperature and particle size were kept constant. This allowed for more widespread recombination of clusters, facilitating a more rapid and representative equilibration of the system. At high dipole moments you get locked and stuck very quickly in metastable states. A lower dipole moment means the free energy landscape is less extreme and metastablity is less prevalent. Furthermore, you could argue that annealing in experiment or simulation would allow one to achieve the same end at higher dipole moments. By reducing the dipole moment we have negated the need for this approach. The magnitude of the dipole moment simply alters the kinetics of the situation but not the final structures, which are of interest here. A further experimental justification of this approach is due to the fact that the spherical particles are magnetic cubes surrounded by a polystyrene shell effectively shielding the dipole moment. In terms of the short-range interaction the value of $r_{\text{off}}$ is set such that the net force between two particles at close contact due to the total interaction potential is zero. Furthermore, the energy parameter was increased to $\epsilon=1000$ to reduce the softness of the interaction. 

Simulations were conducted on systems of $N_{c}=1000$ clusters, in a strictly two dimensional geometry, {\it i.e.} clusters were not permitted to rotate out of plane, only in-plane. Periodic boundary conditions were implemented and dipolar interactions were handled using the P$^{3}$M algorithm in combination with a dipole layer correction, both with an accuracy on the order of $10^{-4}$ in the forces\cite{Cerda:2008jcp,BRODKA200462}. Due to the fixed monolayer geometry of the system, three situations arise in terms of dipole configurations due to the effect of chirality. The first being a system of clusters where the dipole configuration of each cluster circulates in one direction {\it i.e.} anti-clockwise. The second being the antithesis of this, a dipolar configuration circulating in the other direction, {\it i.e.} clockwise. The third option is a mixture of these two geometry-enforced cluster types, we decided to investigate a 50:50 racemic mixture of clockwise and anti-clockwise clusters. 

Simulations were performed in the NVT ensemble, where the system was initialised by randomly placing and rotating the clusters within the plane at an area fraction of $\varphi_{A}=0.4$. The system was then propagated again using Langevin molecular dynamics from this initial configuration for a total of $2.0\times10^{5}\;\Delta t^{*}$, with $\Delta t^{*}=0.001$ as before. Configurations were recorded at intervals of $1.0\times10^{3}\;\Delta t^{*}$ to monitor the evolution of the aggregation. Simulations were again performed using ESPResSo 3.3.0\cite{Arnold:2013esp}. The final recorded configuration was then visualised and feature as the snapshots in the main text and Supporting Information. For the visualisation of the sub-lattices within the aggregate cut-off radii were used to draw the bonds, where $R^{*}_{\text{b}}=1.4$, and $R^{*}_{\text{hc}}=2.1$ for the bounce and honeycomb lattice respectively. For the dipolar (staggered) kagome lattice, the visualisation was created by drawing tangents along the dipole moments. 
\\
\\
%=================
%=================
% Acknowledgments 
%=================
%=================

\acknowledgments
 L.R. acknowledges the Netherlands Organisation for Scientific Research (NWO) for financial support through a VENI grant (680-47-446). S. Kantorovich, S. Schyck, J.M. Meijer, S. Sacanna, T. Huekel, K. Masania and C. Storm are thanked for many valuable discussions. We are grateful to F. Grozema for the use of computational resources at TU Delft.

\providecommand{\latin}[1]{#1}
\makeatletter
\providecommand{\doi}
  {\begingroup\let\do\@makeother\dospecials
  \catcode`\{=1 \catcode`\}=2 \doi@aux}
\providecommand{\doi@aux}[1]{\endgroup\texttt{#1}}
\makeatother
\providecommand*\mcitethebibliography{\thebibliography}
\csname @ifundefined\endcsname{endmcitethebibliography}
  {\let\endmcitethebibliography\endthebibliography}{}

\end{document}

\title{Supporting Information:
	\\Magnetic Coupling in Colloidal Clusters  for Hierarchical Self-Assembly}
\author[1,2]{Joe G. Donaldson}
\author[3]{Peter Schall}
\author[1]{Laura Rossi $^\star$}

\affil[1]{Department of Chemical Engineering, Delft University of Technology, 2629 HZ Delft, The Netherlands}
\affil[2]{Current Address: Unilever R\&D, Colworth MK44 1LQ, UK}
\affil[3]{Institute of Physics, University of Amsterdam, 1098XH Amsterdam, The Netherlands.}

\renewcommand\Authands{ and }

\maketitle

\begin{footnotesize}
	$^\star$      L.Rossi@tudelft.nl
	\\
\end{footnotesize}

\doublespacing

\doublespacing

\newpage
\newpage

	%------------------
% MAIN TEXT
%------------------
\section*{Cluster Property Comparison.}
\begin{figure}[p!]	\renewcommand{\thefigure}{S\arabic{figure}}
	\centering
	\includegraphics[scale=1.00]{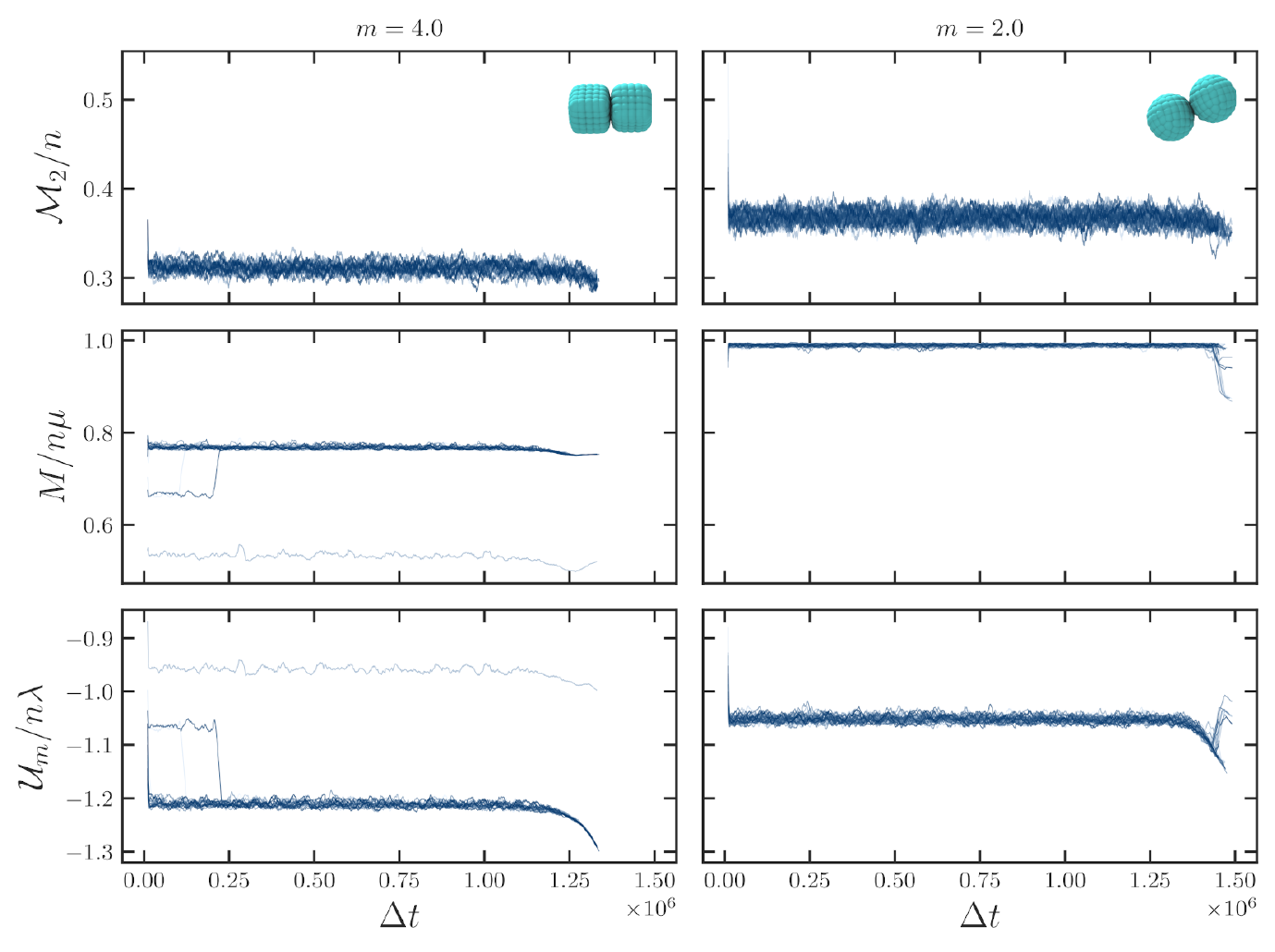}
	\caption{\textbf{Cluster Size:} $n=2$}  
    \label{fig_N2}
\end{figure}

\begin{figure}[htb!]\renewcommand{\thefigure}{S\arabic{figure}}
	\centering
	\includegraphics[scale=1.00]{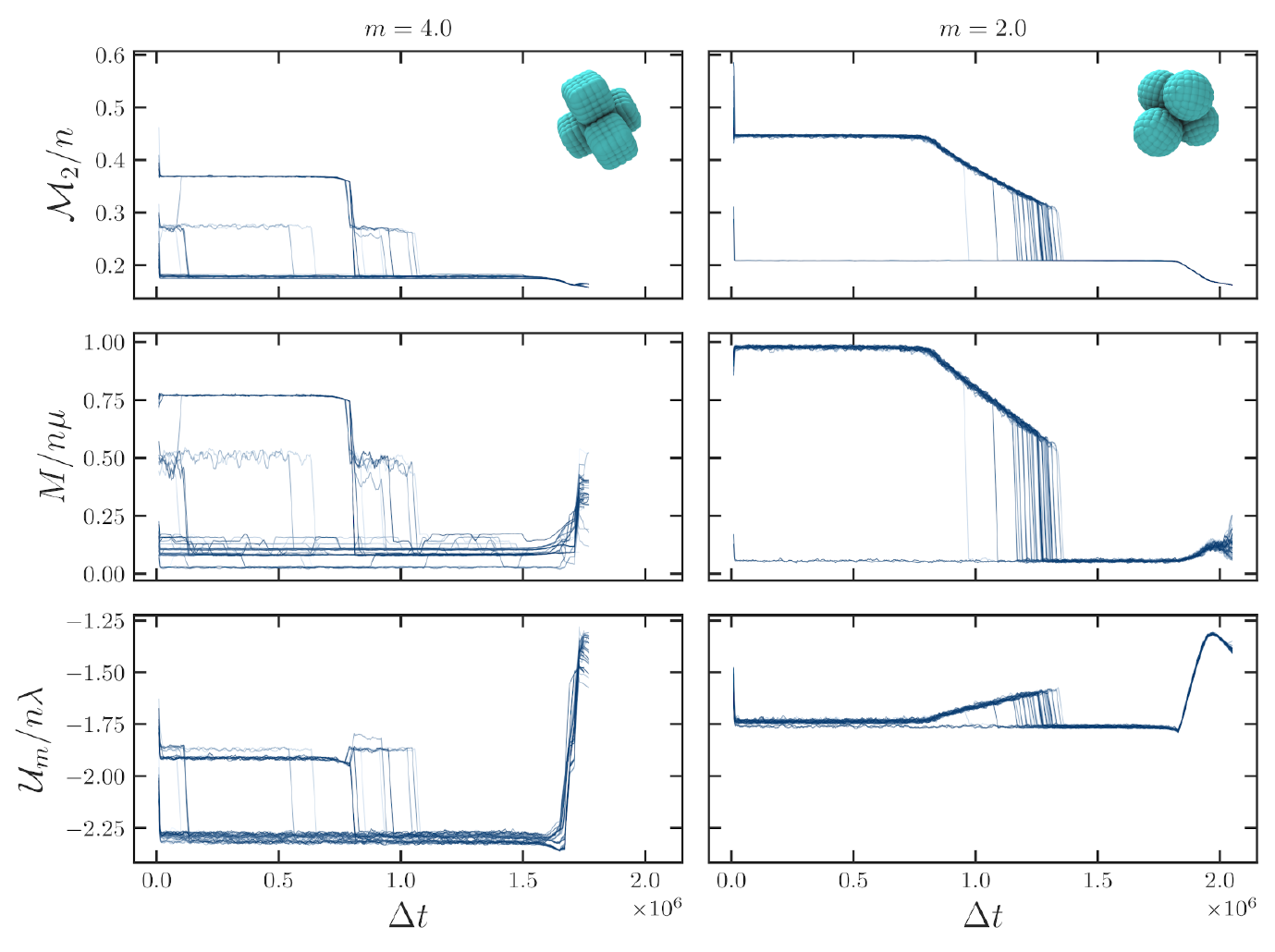}
    \caption{\textbf{Cluster Size:} $n=4$}   
    \label{fig_N4}
\end{figure}

\begin{figure}[htb!]\renewcommand{\thefigure}{S\arabic{figure}}
	\centering
	\includegraphics[scale=1.00]{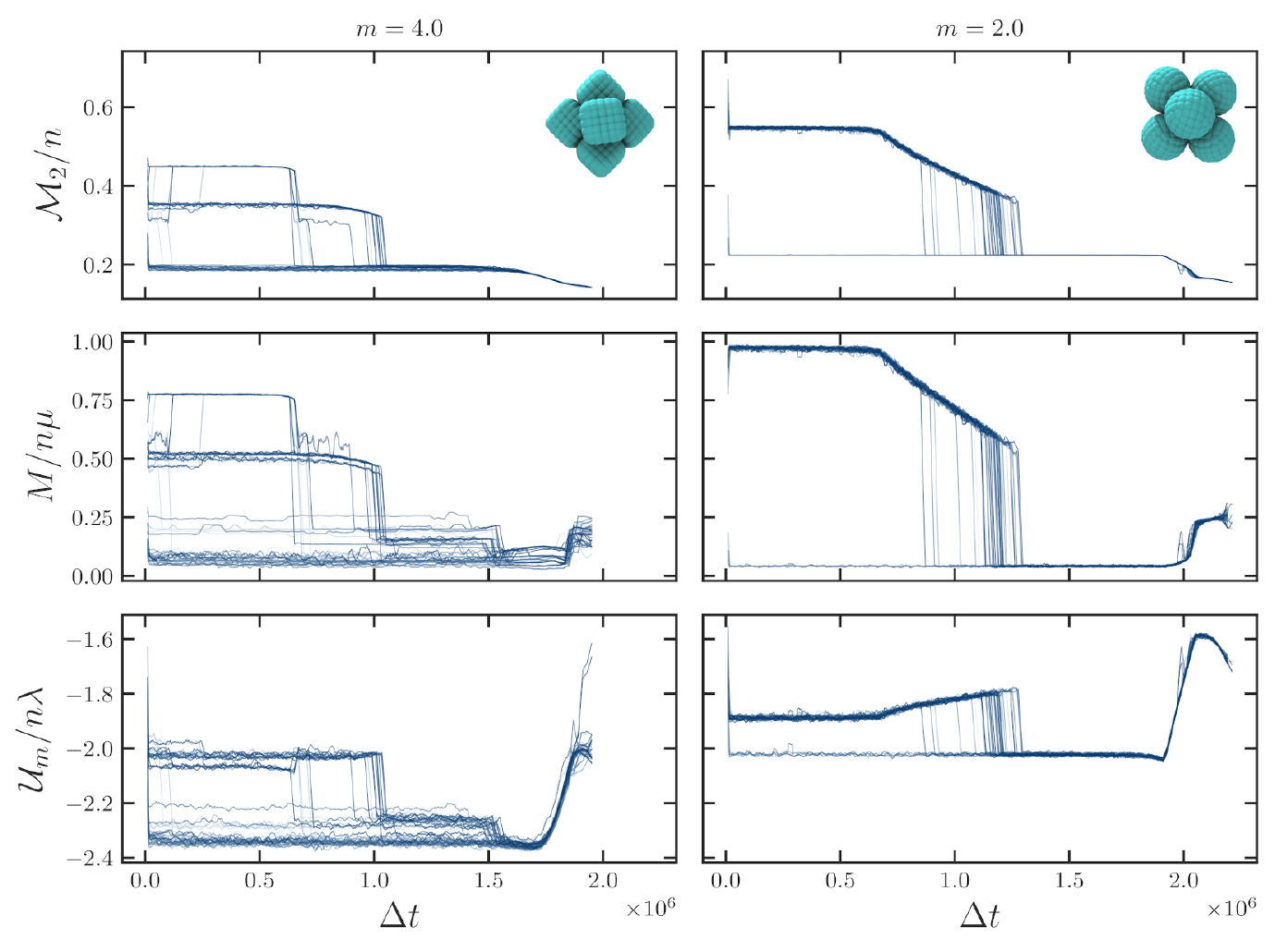}
    \caption{\textbf{Cluster Size:} $n=5$}   
    \label{fig_N5}
\end{figure}

\begin{figure}[htb!]\renewcommand{\thefigure}{S\arabic{figure}}
	\centering
	\includegraphics[scale=1.00]{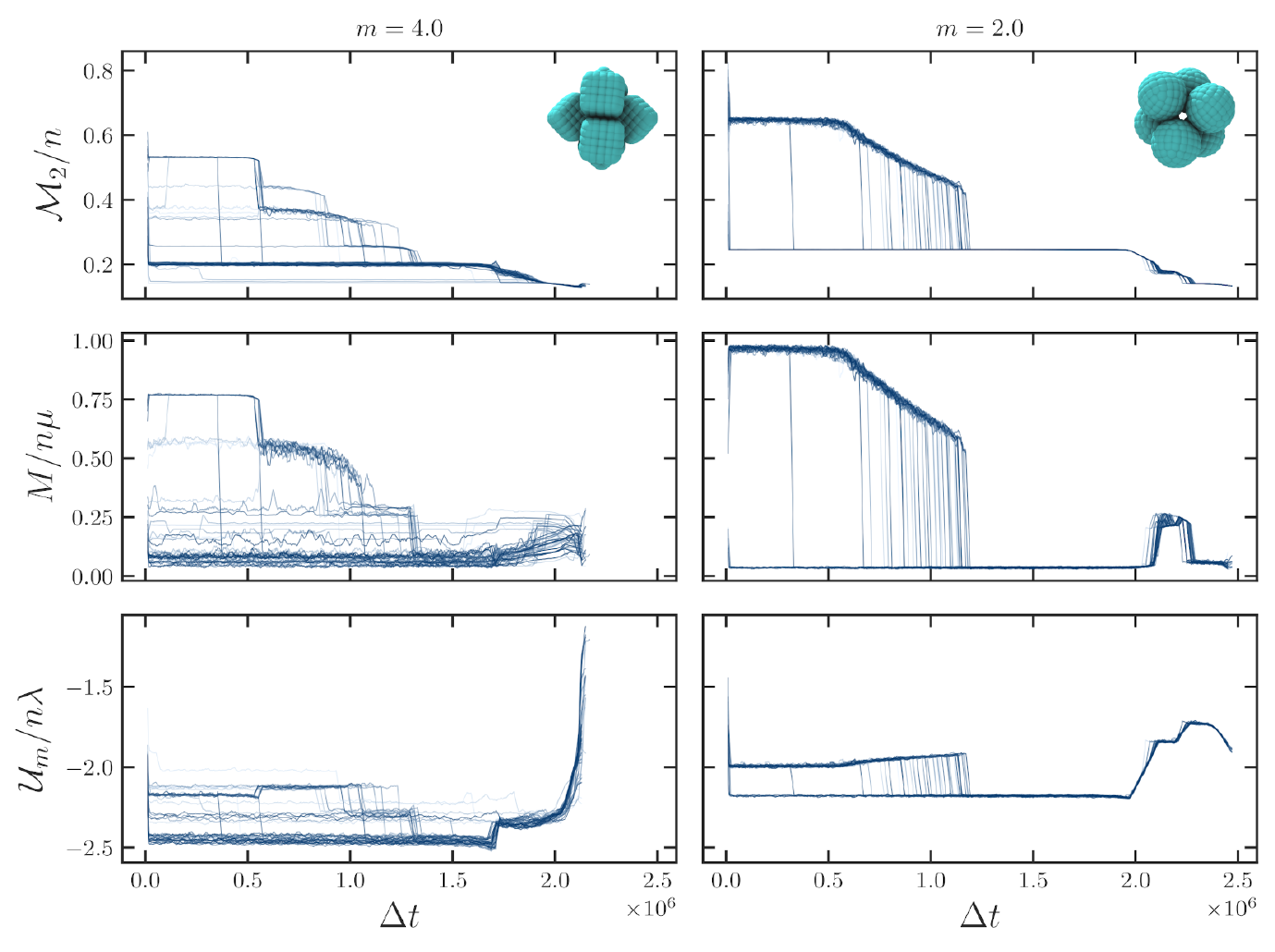}
    \caption{\textbf{Cluster Size:} $n=6$}   
    \label{fig_N6}
\end{figure}

\begin{figure}[htb!]\renewcommand{\thefigure}{S\arabic{figure}}
	\centering
	\includegraphics[scale=1.00]{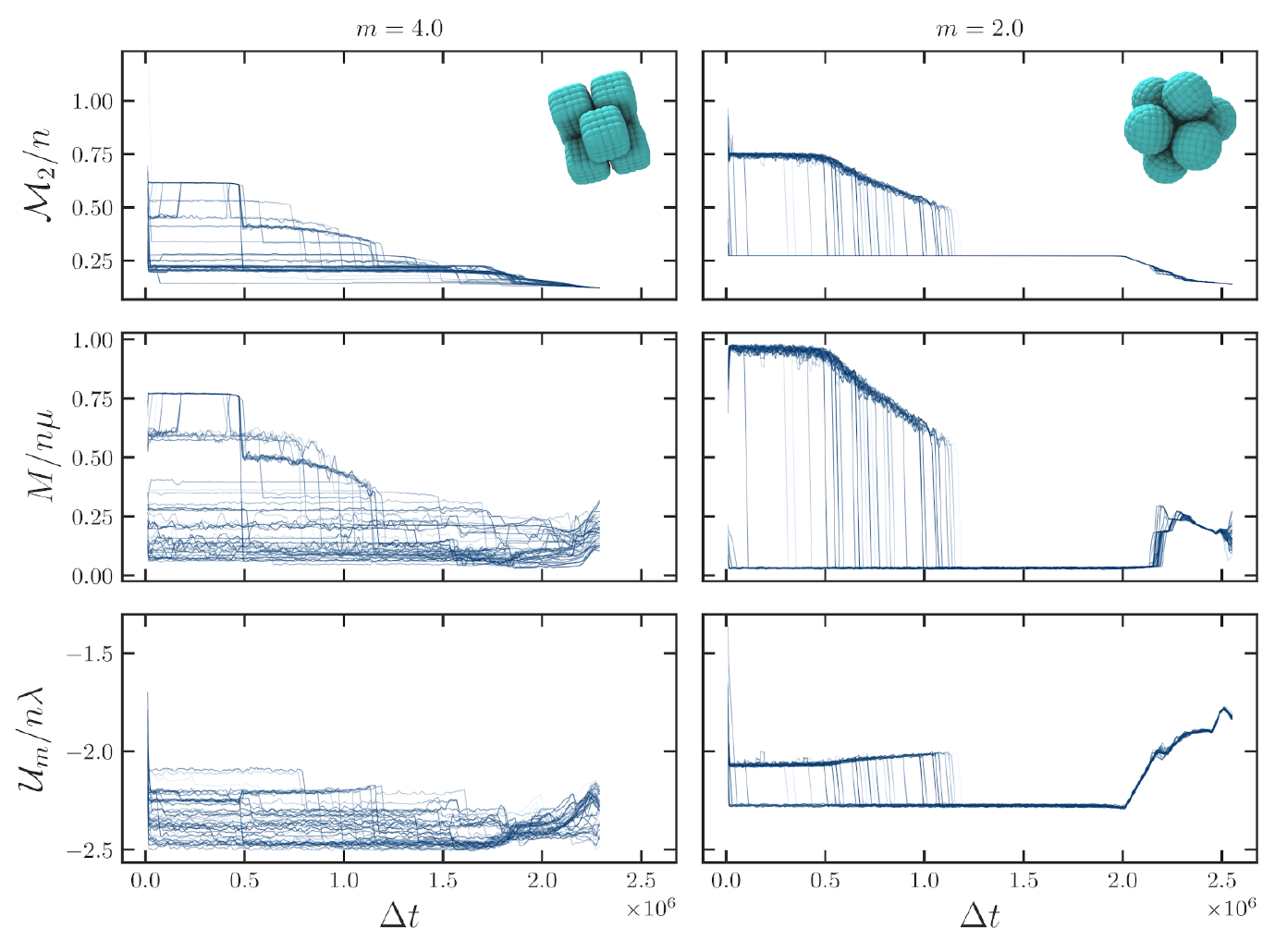}
    \caption{\textbf{Cluster Size:} $n=7$}   
    \label{fig_N7}
\end{figure}

\begin{figure}[htb!]\renewcommand{\thefigure}{S\arabic{figure}}
	\centering
	\includegraphics[scale=1.00]{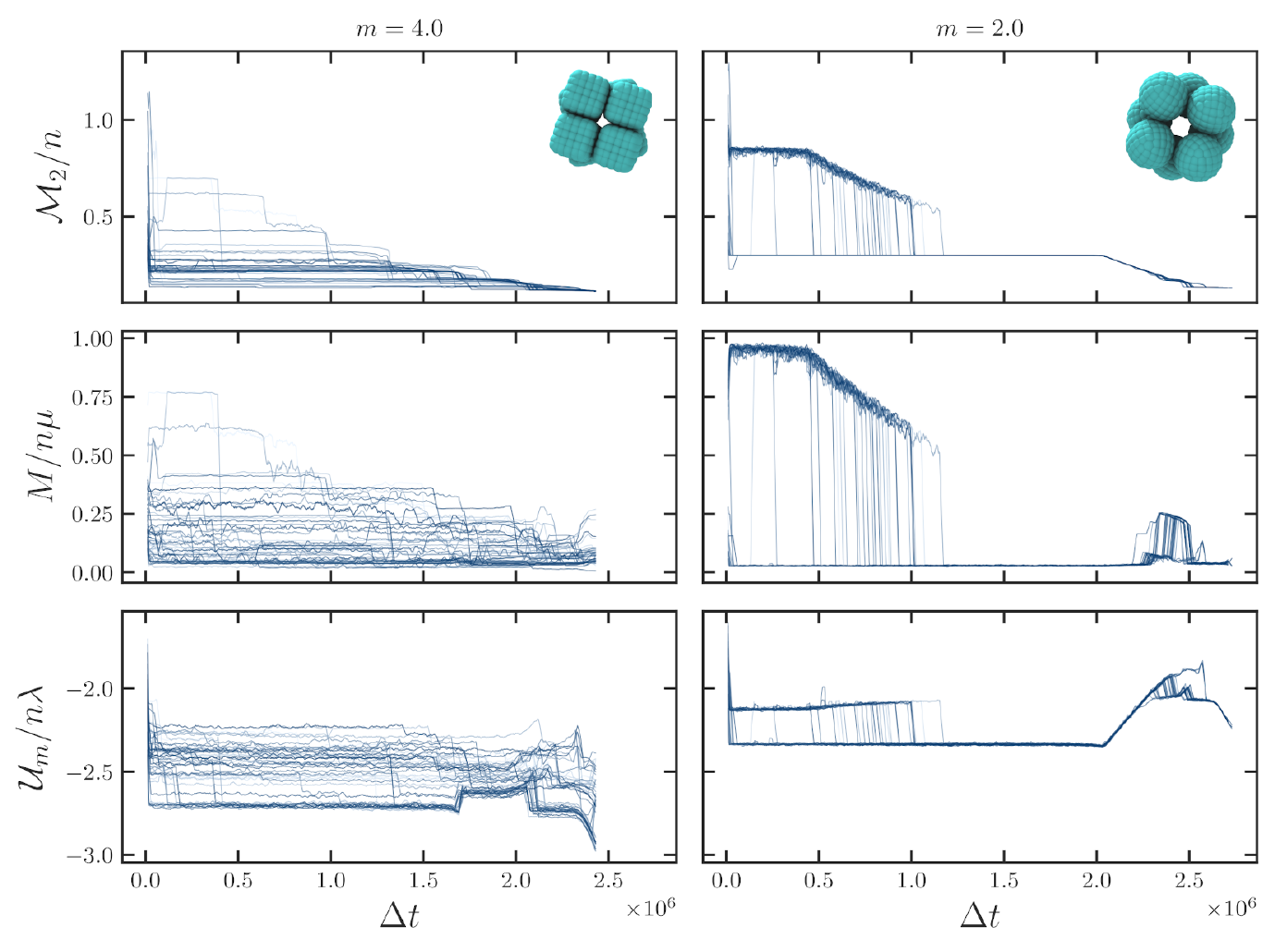}
    \caption{\textbf{Cluster Size:} $n=8$}   
    \label{fig_N8}
\end{figure}

\begin{figure}[htb!]\renewcommand{\thefigure}{S\arabic{figure}}
	\centering
	\includegraphics[scale=1.00]{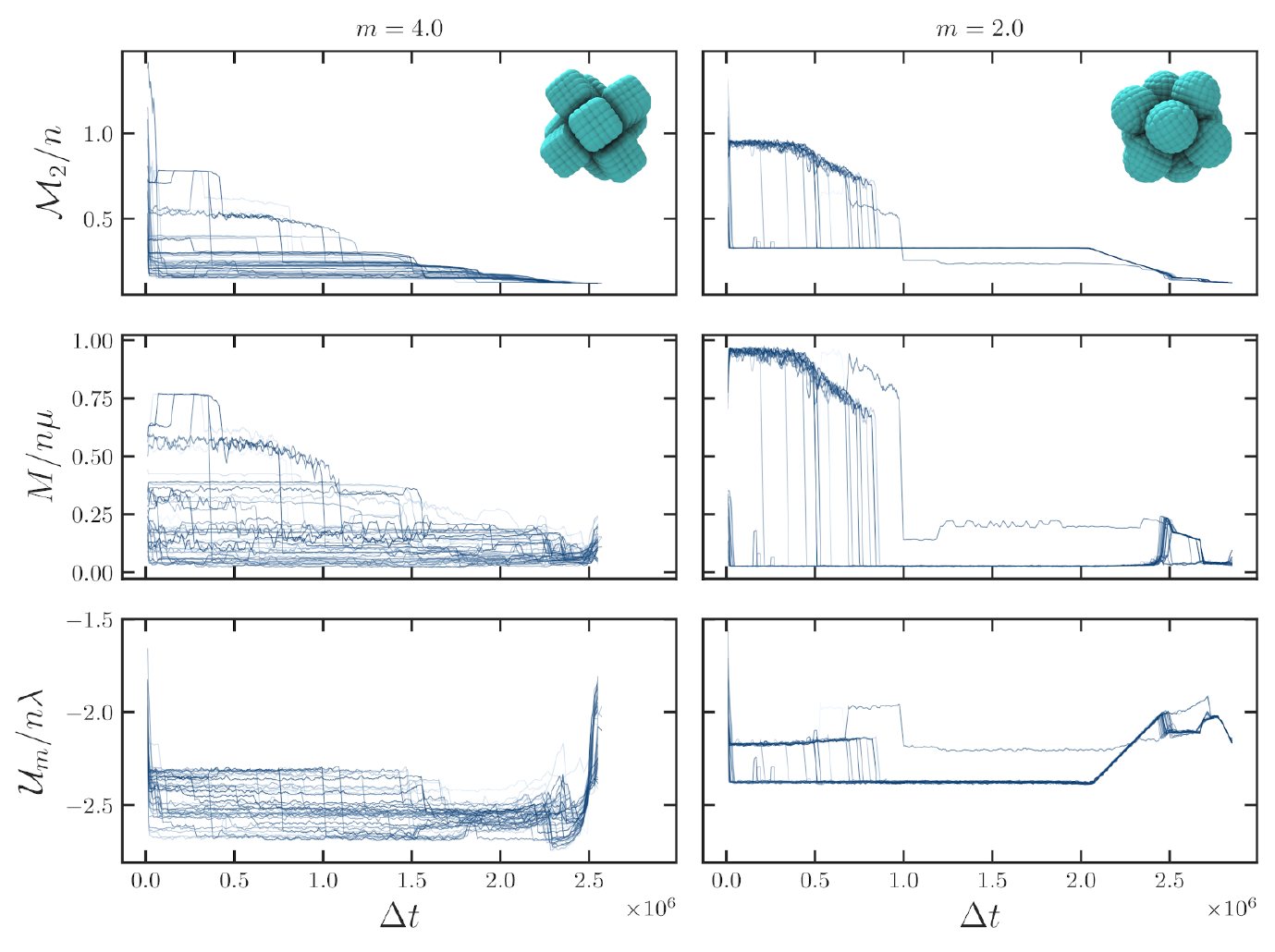}
    \caption{\textbf{Cluster Size:} $n=9$}    
    \label{fig_N9}
\end{figure}

\begin{figure}[htb!]\renewcommand{\thefigure}{S\arabic{figure}}
	\centering
	\includegraphics[scale=1.00]{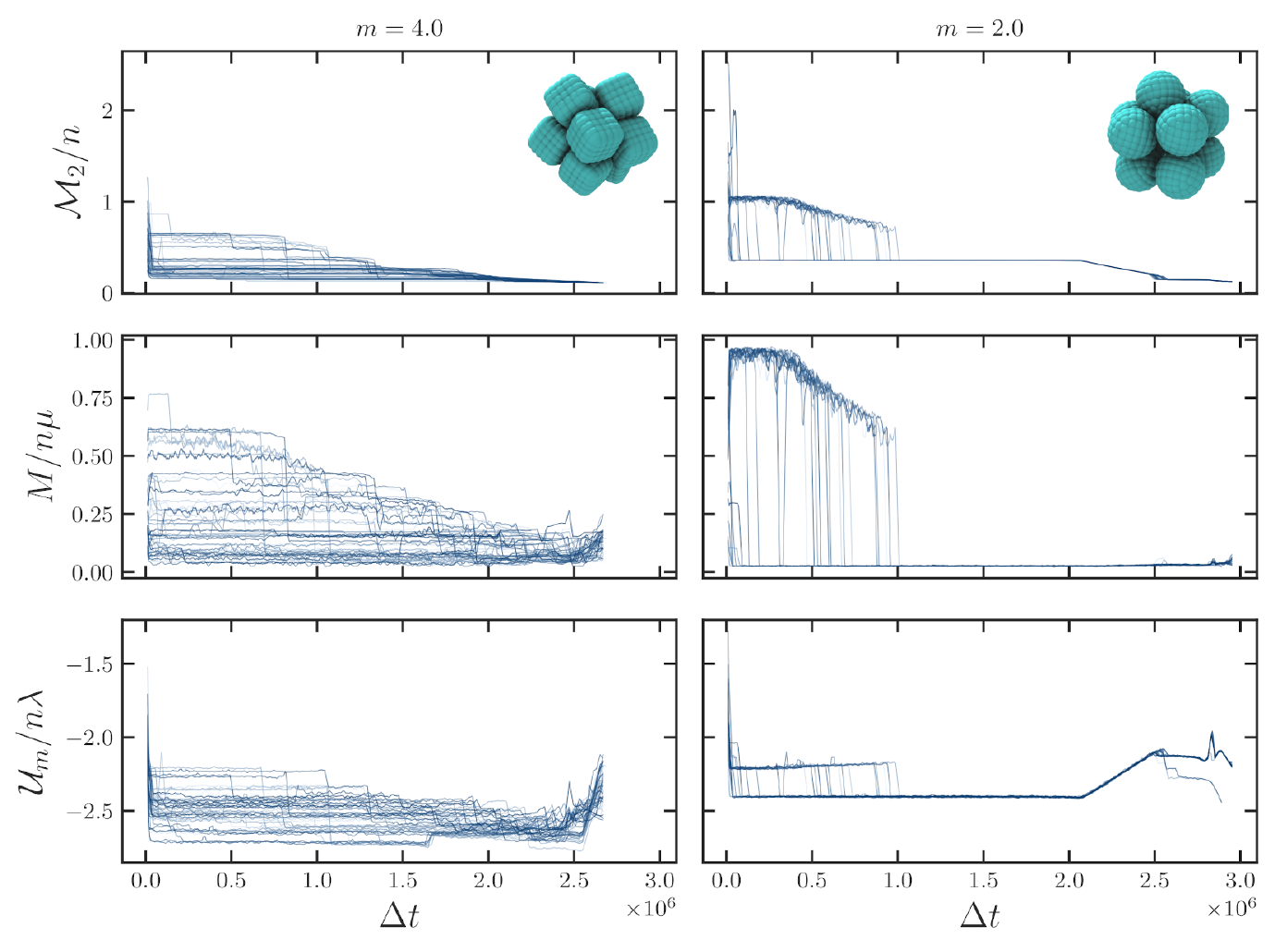}
    \caption{\textbf{Cluster Size:} $n=10$}    
    \label{fig_N10}
\end{figure}

In Figures~\ref{fig_N2}-\ref{fig_N10} we showcase the data for the other cluster sizes investigated in this work and not singularly shown in the main text, namely $n=2,[4,10]$. It is important to recall however that this cluster data is summarised in Figure~5 of the manuscript as well. As a reminder to the reader, Figures~\ref{fig_N2}-\ref{fig_N10} describe and monitor cluster formation for each cluster size respectively and compare the evolution for sphere and cube clusters. The plot grid displays the data for each particle type in each column, cubes ($m=4$)  and spheres ($m=2$). Each cluster property observable is plotted in each row as follows, second moment of the mass distribution $\mathcal{M}_{2}$, total dipole moment of a cluster  $M$, and the total magnetic interaction energy $\mathcal{U}_{m}$. Each quantity is normalised to allow the data for different cluster sizes to be viewed on an equal footing. The evolution of each quantity is plotted in units of the simulation time-step $\Delta t$. The fifty replica compression runs performed for each type of cluster are shown in each plot, provided that the run completed successfully. The evolution of each replica was smoothed by calculating the moving average over 200 measures. The compression scheme used meant that the reduction in droplet size at each iteration was reduced as the simulation progressed. As the iteration length was kept fixed this equates to the confinement being applied more slowly as the system increased in density and thus facilitated the search for equilibrated structures.

The same conclusions reached in the manuscript with regards to the similarities and differences in cluster formation for cubes and spheres are applicable for the cluster sizes shown here, save for a few minor points which we will address shortly. The conclusion of the manuscript can be summarised as the ability of sphere clusters to repeatedly reproduce the same structural arrangements of the constituent particles but also their dipole configurations as well. Cubes struggle to achieve both, managing the former, not the latter. This summary holds true for n=$[4,9]$, where we see a convergence of $\mathcal{M}_{2}$, $M$, and $\mathcal{U}_{m}$ for all replicas in the case of spherical particles, but multiple final values in $M$, and $\mathcal{U}_{m}$ but not  $\mathcal{M}_{2}$ for cube clusters indicating the structural consistency but magnetic frustration. In the case of $n=2$ we have some broader variations in the terminal values of $M$, and $\mathcal{U}_{m}$ in the replicas, this can be attributed to the fact that, as the evaporation process comes to an end, the linearity of the structure makes it less rigid and therefore more susceptible to deformations by the evaporating droplet. A similar observation can be made for the spherical cluster of $n=10$, where $\mathcal{U}_{m}$ has a bifurcation, a sign of two structures with different dipolar configurations. We attribute this to the fact that as the cluster size increases, the number of available microstates increases rapidly, making it harder for the system to consistently rearrange to the same final state, it seems the $n=10$ is where this effect begins to manifest.  

%------------------
% FIGURE N=2
%------------------

%------------------
% FIGURE N=4
%------------------

%------------------
% FIGURE N=5
%------------------

%------------------
% FIGURE N=6
%------------------

%------------------
% FIGURE N=7
%------------------

%------------------
% FIGURE N=8
%------------------

%------------------
% FIGURE N=9
%------------------

%------------------
% FIGURE N=10
%------------------

\section*{Cluster Aggregation}

\begin{figure}[htb!]\renewcommand{\thefigure}{S\arabic{figure}}
	\centering
	\includegraphics[scale=0.85]{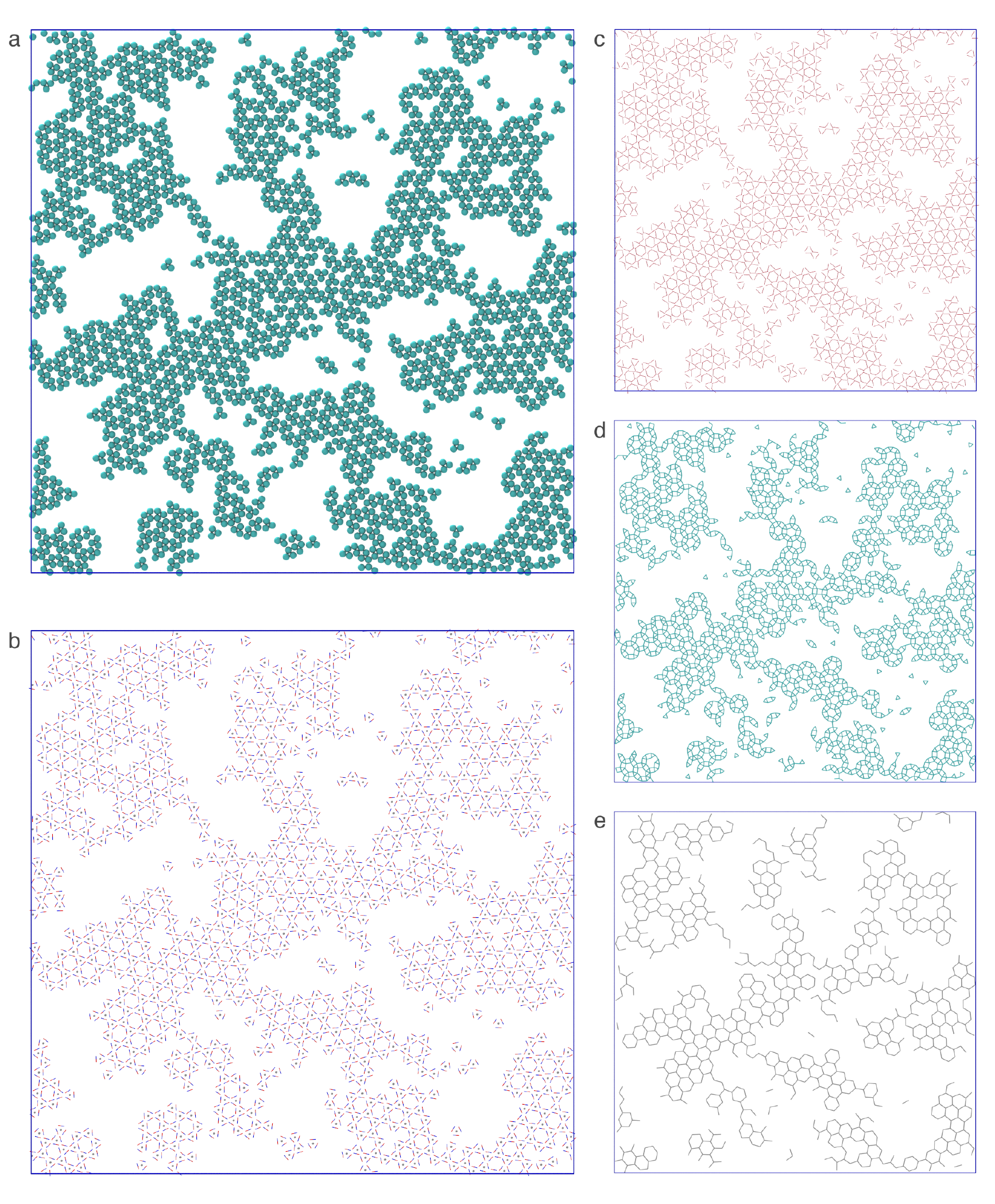}
	\caption{\textbf{ Monolayer Snapshot:} Clockwise trimers only at $\varphi_{A}=0.4$ (a) Monolayer structure. (b) Dipole configuration, where the centre of mass of each cluster is indicated by the sliver sphere. (c) Dipole lattice with staggered kagome symmetry (see main text for details about the structure). (d) Particle lattice with bounce symmetry. (e) Lattice based on the cluster centre of mass with honeycomb symmetry. }  
    \label{fig_clock}
\end{figure}

\begin{figure}[htb!]\renewcommand{\thefigure}{S\arabic{figure}}
	\centering
	\includegraphics[scale=0.85]{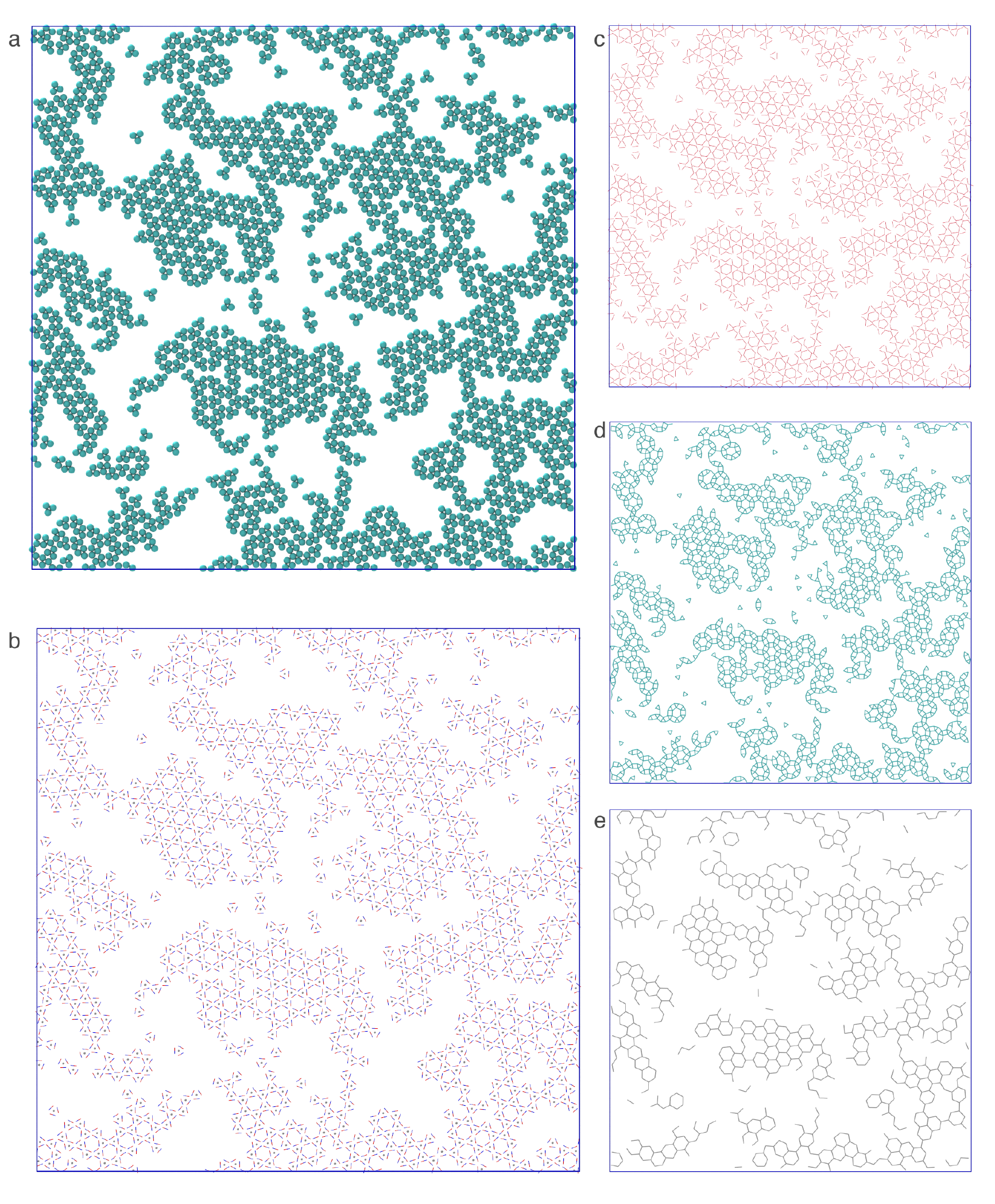}
    \caption{\textbf{ Monolayer Snapshot:} Anticlockwise trimers only at $\varphi_{A}=0.4$ (a) Monolayer structure. (b) Dipole configuration, where the centre of mass of each cluster is indicated by the sliver sphere. (c) Dipole lattice with staggered kagome symmetry (see main text for details about the structure). (d) Particle lattice with bounce symmetry. (e) Lattice based on the cluster centre of mass with honeycomb symmetry. }  
    \label{fig_anticlock}
\end{figure}

\begin{figure}[htb!]\renewcommand{\thefigure}{S\arabic{figure}}
	\centering
	\includegraphics[scale=0.85]{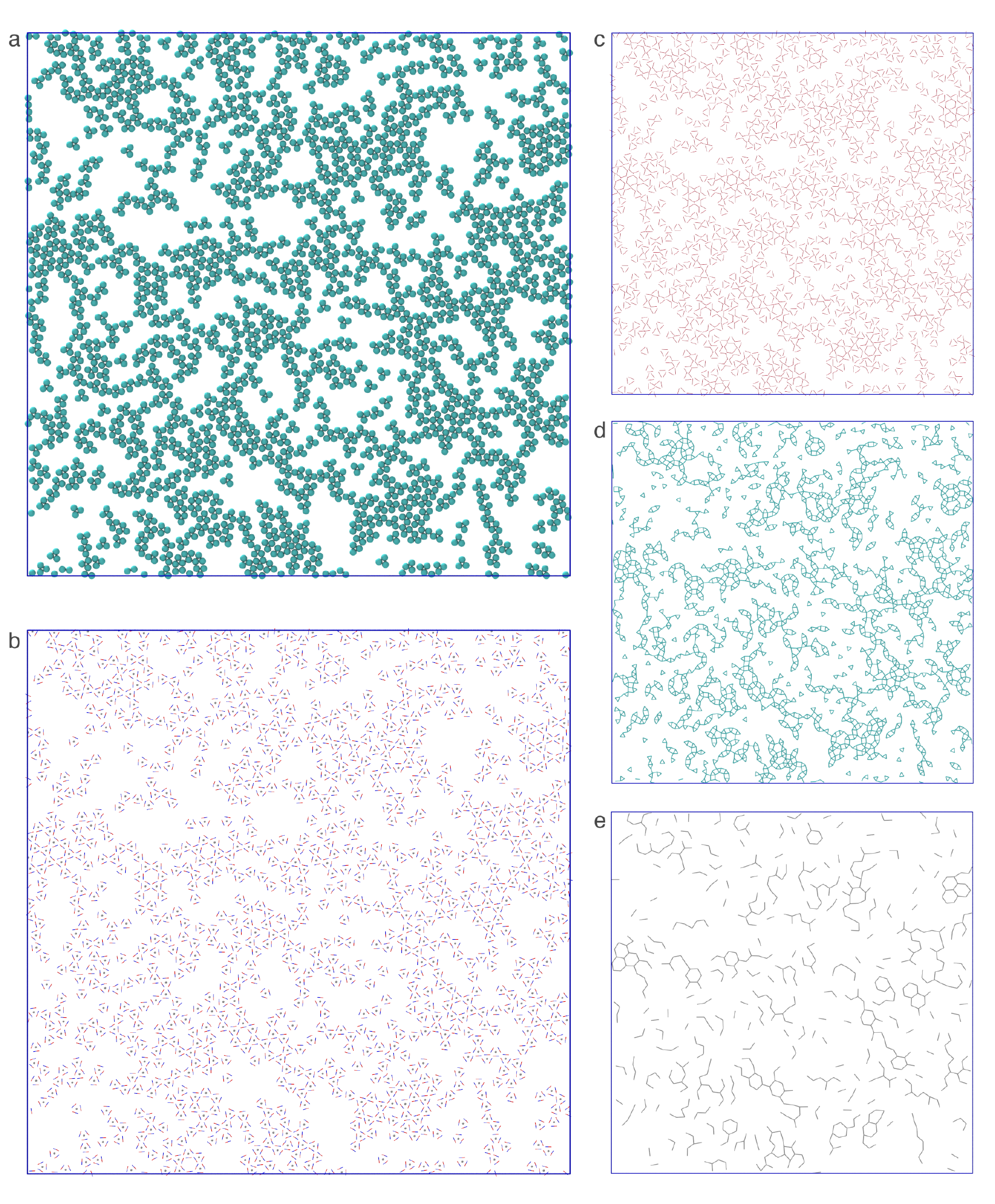}
    \caption{\textbf{ Monolayer Snapshot:} Racemic mixture of trimers at $\varphi_{A}=0.4$ (a) Monolayer structure. (b) Dipole configuration, where the centre of mass of each cluster is indicated by the sliver sphere. (c) Dipole `lattice'.(d) Particle `lattice'. (e) Lattice based on the cluster centre of mass. (c-e) Lattice formation here is extremely limited and only occurs in isolated regions and has minimal extent.}    
    \label{fig_race}
\end{figure}

In this section we reproduce the monolayer snapshots in the full field of view for the clockwise case. This is shown in Figure~\ref{fig_clock}. In contrast to the manuscript, here in Figure~\ref{fig_clock}c-e we visualise the lattice pattern only and do not show the gradient transition. This choice holds for the subsequent figures as well. In Figure~\ref{fig_anticlock} we show the monolayer for the anticlockwise oriented cluster. As can be seen from the stills, the structure of the aggregated monolayer is akin to the clockwise variant as one would expect. However, moving to Figure~\ref{fig_race} where we visualise the racemic mixture, one clearly notes that the monolayer is significantly less aggregated. The presence of the two enantiomers frustrates the self-assembly process resulting only in small areas of agglomeration between clusters of the same handedness. Given sufficient time one would expect the system to perform a kind of phase separation into distinct regions of each enantiomer. It is clear that when chirality is imposed on the system due to the strict two dimensional topology of the monolayer that enatiomerically pure systems offer the best route to hierarchical assembly. 
%------------------
% FIGURE RIGHT
%------------------

%------------------
% FIGURE LEFT
%------------------

%------------------
% FIGURE RACEMIC
%------------------
